\documentclass[10pt,journal,final]{IEEEtran}
\usepackage{amsmath,amsfonts,amsthm,amssymb,bm}
\usepackage{algorithmic,algorithm,array}
\usepackage[caption=false,font=small,labelfont=rm,textfont=rm]{subfig}
\usepackage{textcomp}
\usepackage{stfloats}
\usepackage{url}
\usepackage{verbatim}
\usepackage{graphicx}
\usepackage{cite,threeparttable}
\usepackage{diagbox, multirow, makecell, booktabs}
\usepackage{xcolor, balance}

\newtheorem{Problem}{\textbf{Problem}}
\newtheorem{Definition}{Definition}
\newtheorem{Proposition}{Proposition}
\newcounter{MYtempeqncnt}
\theoremstyle{remark}

\begin{document}

\title{Multimodal Semantic Communication using Shared and Unique Representation Decomposition}

\author{\IEEEauthorblockN{Peng Yi,~\IEEEmembership{Graduate Student Member, IEEE}, and Ying-Chang Liang,~\IEEEmembership{Fellow, IEEE}}\\ 
\thanks{\par Peng Yi is with the National Key Laboratory of Wireless Communications, and the Center for Intelligent Networking and Communications (CINC), University of Electronic Science and Technology of China, Chengdu 611731, China (email: yipengcd@outlook.com).
\par Ying-Chang Liang is with the Institute of Fundamental and Frontier Sciences and the Center for Intelligent Networking and Communications (CINC), University of Electronic Science and Technology of China (UESTC), Chengdu 611731, China (email: liangyc@ieee.org). The corresponding author is Y.-C. Liang.
}}

\maketitle
\begin{abstract}
    Semantic communication (SemCom) has emerged as a promising paradigm for bandwidth-constrained mobile and vehicular networks. This paper proposes a unified representation learning framework for multimodal SemCom (MSC), which encodes heterogeneous data into a compact semantic space for efficient transmission and source recovery. The framework addresses the multimodal semantic representation gap, which arises when at least one modality contains semantic information that cannot be fully inferred from the other. To preserve this information for reconstruction, the framework decomposes each representation into shared and modality-specific components. We formulate a mutual information objective for source reconstruction, cross-modal alignment of the shared components, and decoupling of the shared and modality-specific components. To enhance transmission efficiency, we introduce a shared knowledge base implemented as a compact codebook of semantic vectors learned during training. The shared components are transmitted digitally as codebook indices, while the modality-specific components are sent in analog form. We apply knowledge distillation during training, using a pretrained vision foundation model to guide the learning of quantized shared features. On image--text pairs over additive white Gaussian noise (AWGN) and Extended Vehicular A (EVA) channels, the proposed framework improves image reconstruction over conventional separate source-channel coding (SSCC) and representative multimodal and unimodal SemCom systems. Further evaluations show gradual degradation under channel-estimation mismatch, complementary effects of alignment, decoupling, and knowledge distillation on the shared representation, and alignment of shared features across modalities with separation of the modality-specific features.
    \end{abstract}


\section{Introduction}
Next-generation wireless systems increasingly integrate artificial intelligence (AI) with communication technologies~\cite{chinacom19/lin}, owing to the growing computational capability of edge devices and the widespread deployment of AI techniques~\cite{jsac22/khaled}. In this setting, semantic communication (SemCom) has emerged as an approach to AI-native communications~\cite{comst26/loc}. Whereas conventional bit-centric systems aim to recover the transmitted bits, SemCom prioritizes the intended meaning of a message. By extracting task-relevant semantic information while suppressing redundancy, SemCom can reduce transmitted data volume and improve robustness to noise. These advantages matter for vehicular and mobile links, where spectrum is scarce and channels vary rapidly.

Given that SemCom systems must handle diverse data, differences in storage formats and in how information is embedded call for distinct system designs. For instance, convolutional neural networks (CNNs)~\cite{tccn19/eirina, tvt25/xiaoyi} and Swin Transformers~\cite{tccn25/ke, tvt24/loc} are typically used for images and point clouds, whereas the vanilla Transformer~\cite{tsp21/huiqiang, tvt25/mengli} is commonly used for text. Accordingly, many SemCom systems target a single modality by exploiting its structural and semantic properties~\cite{comst24/zhilin}. However, emerging vehicular applications, including in-vehicle multimedia and vehicle-to-everything (V2X) information sharing, increasingly require joint transmission of multimodal data such as images, text, and audio~\cite{tvt26/tian}. Existing SemCom systems, mostly designed for unimodal data, are poorly suited to these multimodal settings. Transmitting multimodal data with these systems requires multiple unimodal frameworks, which adds processing overhead and leaves inter-modal semantic correlations unused.

To address these challenges, multimodal SemCom (MSC) has emerged as an integrated framework for joint processing of multimodal data. Early MSC work focused mainly on multimodal fusion, in which heterogeneous modalities are integrated for joint tasks. This paradigm is termed multimodal understanding SemCom. 
A representative example is visual question answering (VQA)~\cite{jsac22/huiqiang}, where text and image inputs are encoded by parallel deep neural networks (DNNs) at the transmitter. After per-modality semantic decoding, the receiver performs feature-level fusion to produce textual answers.
Building on inter-modal dependency, video conferencing systems in~\cite{jsac23/pulkit, wcl25/haonan} exploit strong audio--visual correlation to reconstruct part of the video from the audio stream, thereby cutting bandwidth while preserving contextual fidelity.
To further improve spectral efficiency, over-the-air computation (AirComp) has been introduced for over-the-air feature fusion in SemCom~\cite{tcom25/peng, wc24/xuewen, tvt25/shuangying}. Unlike sequential transmission, AirComp exploits the superposition of wireless signals at the physical layer, enabling simultaneous transmission and joint processing of multimodal features.

\subsection{Motivation}
Despite recent advancements in MSC, multimodal representation learning remains a critical and underexplored frontier in SemCom. 
This process maps heterogeneous data into a unified semantic space and enforces cross-modal alignment for coherent understanding.
For multimodal recovery SemCom, the unified representation must preserve enough information to reconstruct each original modality, rather than retaining only task-specific semantics.
An early effort toward this goal~\cite{tcom24/guangyi} proposed a multi-task learning-based MSC framework, where a shared semantic decoder reconstructs multiple modalities and demonstrates the feasibility of joint representation learning.
However, the modality-specific codebooks partition semantic features into separate subspaces, which limits unified cross-modal representation learning. 
To address this shortcoming, \cite{cm25/feibo} leveraged multimodal large language models (LLMs) to map multimodal inputs into textual representations compatible with text-centric SemCom systems.
While this approach improves interoperability, intensive LLM processing incurs prohibitive computational cost. 
More fundamentally, text-centric encoding can discard non-linguistic semantics, such as spatial structure in images and prosodic cues in speech, which weakens multimodal recovery and limits the generality of text-centric representations.

In this paper, we propose an MSC framework centered on multimodal recovery. The framework aligns shared semantics in a common representation space and preserves modality-specific details for reconstruction. The learned representations are designed to be compact and robust to channel noise.
This design poses three challenges.
\begin{itemize}
    \item \textbf{Challenge 1} \textit{(Heterogeneous Continuous--Discrete Modalities):}
    Continuous sources (e.g., images) and discrete sources (e.g., text) differ in structure and statistics, making it difficult to support both within one SemCom architecture~\cite{tvt26/xuefei}.
    \item \textbf{Challenge 2} \textit{(Multimodal Semantic Representation Gap):}
    One modality may contain semantic information that cannot be fully inferred from the other. The system must transmit shared semantics efficiently while preserving modality-specific details for source recovery.
    \item \textbf{Challenge 3} \textit{(Wireless Signal Distortion):}
    Channel fading, additive noise, and imperfect channel state information (CSI) can degrade the received representations. Errors in either the shared or the modality-specific components can impair source recovery.
\end{itemize}

\subsection{Our Contributions}
To address these challenges, we develop an MSC framework that learns unified representations of heterogeneous data for reliable transmission. The major contributions of this work are summarized as follows. 

\begin{itemize}
    \item[(\romannumeral1)] \textbf{Unified MSC Framework:} We propose an MSC framework for both continuous (e.g., images) and discrete (e.g., text) data, as illustrated in Fig.\,\ref{fig:framework}. Modality-specific encoders and a unified decoder address \textbf{Challenge 1}. Shared/unique decomposition preserves cross-modal commonality without discarding modality-specific content, addressing \textbf{Challenge 2}. We further show that the cross-modal information left in the unique representations is upper-bounded by the decoupling term plus the alignment gap, thereby linking the training objective to the multimodal semantic representation gap. Training with channel noise and estimation errors mitigates \textbf{Challenge 3}.
    
    \item[(\romannumeral2)] \textbf{Shared Knowledge Base with Hybrid Transmission:} We introduce a shared knowledge base, common to both modalities, that quantizes shared components into compact indices. With unequal protection, shared indices are transmitted digitally, while unique features are sent by analog joint source-channel coding (JSCC), reducing channel uses for recurring cross-modal semantics, while leaving high-dimensional residuals in analog form to limit digital cliffs.

    \item[(\romannumeral3)] \textbf{Knowledge Distillation for Enhanced Recovery:} A pretrained vision foundation model guides the semantic encoder during training to enrich shared visual features for cross-modal alignment, and is not used at inference.

    \item[(\romannumeral4)] \textbf{Experimental Validation:} We compare the proposed system with separate source-channel coding (SSCC) and representative multimodal and unimodal SemCom baselines. The proposed MSC improves multimodal reconstruction across the tested signal-to-noise ratios (SNRs) over additive white Gaussian noise (AWGN) and Extended Vehicular A (EVA) channels. Ablation studies and feature visualizations further examine the learned shared and unique representations.
\end{itemize}

\subsection{Organization and Notation}
Section\,\ref{Sec.Related Work} reviews related work on multimodal representation and mutual information estimation. Section\,\ref{Sec.System Model} presents the system model and formulates the problem in terms of mutual information. Sections\,\ref{Sec.Knowledge Base} and\,\ref{Sec.Knowledge Distillation Aided Multimodal Recovery} describe the shared knowledge base and the knowledge distillation mechanism, respectively. Section\,\ref{Sec.System Implementation} describes the network architecture, objective functions, and training algorithm. Section\,\ref{Sec.Simulations and Results} presents the simulation results, and Section\,\ref{Sec.Conclusion} concludes the paper.

\textit{Notation:} Throughout this paper, boldface uppercase letters denote random variables, while boldface lowercase letters denote vectors, matrices, or tensors, including realizations of random variables. The symbols \(\mathbb{R}\) and \(\mathbb{C}\) denote the sets of real and complex numbers, respectively, and \(\mathbb{R}^{m \times n}\) denotes the set of real \(m \times n\) matrices. The symbols \(\mathbb{E}[\cdot]\), \(\exp(\cdot)\), and \(\log(\cdot)\) denote expectation, the exponential function, and the logarithm, respectively. The notation \(x \sim \mathcal{CN}(\mu,\sigma^2)\) indicates that \(x\) is a circularly symmetric complex Gaussian (CSCG) random variable with mean \(\mu\) and variance \(\sigma^2\).

\section{Related Work}
\label{Sec.Related Work}
\subsection{Unified Multimodal Representation}
The objective of unified multimodal representation is to transform information from diverse modalities into a unified representation space, facilitating cross-modal interaction, fusion, and generation~\cite{csur24/paul}. 
This objective is typically pursued by aligning modalities under constraints in a common latent space, or by learning a modality-agnostic network that extracts information across modalities.
Specifically, CLIP-based methods~\cite{icml21/alec} create a shared latent representation space under the assumption of image--text consistency, leveraging contrastive learning to achieve remarkable zero-shot performance. Nonetheless, ensuring complete consistency of semantic information between the paired modalities remains a challenge~\cite{nips23/yan}. When the semantic content of the image and text is not aligned, mismatched information tends to be discarded, which hinders recovering the sources from the latent space.
On the other hand, current multimodal LLMs based on a decoder-only architecture use a modality-agnostic network to extract information from multiple modalities~\cite{iclr25/chunting}. While these models excel in generating output sequences from latent representations, they lack a mechanism for directly preserving or reconstructing detailed semantic content across modalities. In the context of semantic communication, where accurate reconstruction and the preservation of meaning are critical, this limitation is especially pronounced. Moreover, the latent space in such models is implicit, making it difficult to directly apply them in the channel transmission process inherent to semantic communication.
Factorized representation learning~\cite{nips23/paul} separates shared and unique information, but it targets downstream prediction rather than source recovery. In contrast, the proposed MSC framework retains both components for source recovery and maps them to a compact representation space for noisy-channel transmission.

\subsection{Mutual Information Estimation}
Mutual information is widely recognized as a pivotal metric for quantifying the statistical dependence between two random variables. However, exact mutual information computation is generally intractable except in restricted settings. This limitation arises from the necessity of having closed-form expressions for the associated probability density functions and a tractable formulation of the log-density ratio between the joint and marginal distributions. 
In many deep learning applications, only samples drawn from the joint distribution are available. Consequently, estimators based on deep neural networks have been developed to approximate mutual information from such sample data~\cite{icml19/ben}. Existing mutual information estimators may be broadly categorized into two primary approaches based on their underlying objectives: maximization and minimization. 
In approaches centered on mutual information maximization, the objective is either to extract shared information between random variables, a goal common in representation learning, or to enhance the effective transfer of relevant information between inputs and outputs, as seen in information distillation. This process is typically accomplished by constructing and maximizing a lower-bound estimator, e.g., the mutual information neural estimator (MINE)~\cite{icml18/mohamed}. 
Conversely, applications such as disentangled representation learning and domain adaptation necessitate strategies aimed at minimizing mutual information. In these cases, methodologies like the contrastive log-ratio upper bound (CLUB)~\cite{icml20/pengyu} have been employed to approximate and subsequently minimize the mutual information upper bound, thereby facilitating the disentanglement of modality-specific features from shared ones. 
Building on these estimators, this paper not only maximizes cross-modal mutual information for shared features but also minimizes it between shared and unique features to encourage decoupling.

\begin{figure*}[htbp]
    \centering
    \includegraphics[scale=0.55]{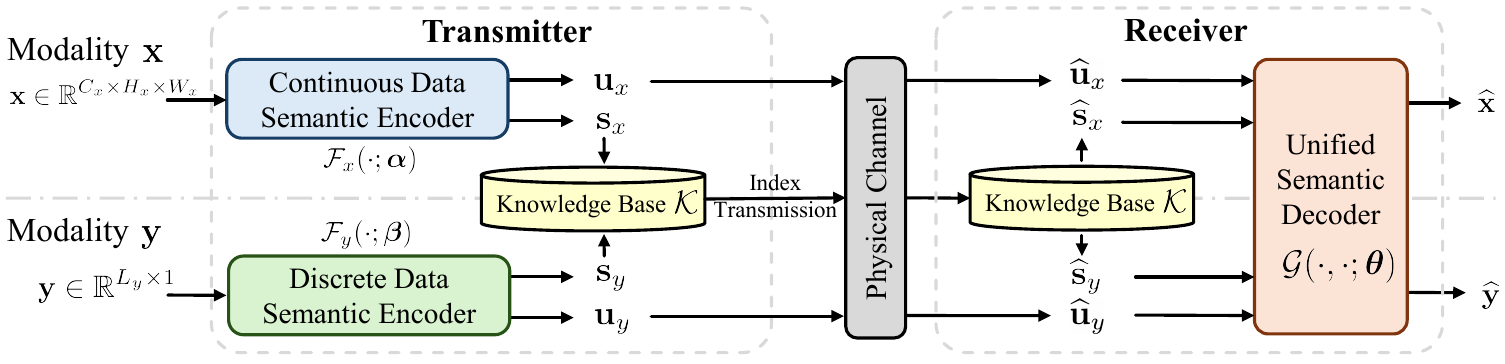}
    \caption{An overview of the proposed MSC framework for continuous (e.g., image) and discrete (e.g., text) data. Each modality is encoded into shared and unique components. Shared components are quantized by a common knowledge base and transmitted as digitally protected indices, while unique components are sent by analog JSCC. Although both modalities are drawn for clarity, each modality can be transmitted and decoded independently. A unified decoder reconstructs the source from the received shared and unique features.}
    \label{fig:framework}
\end{figure*}

\section{System Model and Problem Formulation}
\label{Sec.System Model}
\subsection{The Multimodal Semantic Representation Gap}
\label{Subsec.Hypothesis}
This paper considers two data types: continuous and discrete~\cite{csur24/paul}. Continuous data, typified by image data, are represented by the random variable \(\mathbf{X}\), with realizations \(\mathbf{x} \in \mathbb{R}^{C_x \times H_x \times W_x}\), where \(C_x\), \(H_x\), and \(W_x\) denote the number of channels, height, and width, respectively. Discrete data, analogous to textual information, are represented by the random variable \(\mathbf{Y}\), with realizations \(\mathbf{y} \in \{1, \ldots, V\}^{L_y\times 1} \), where \(V\) is the vocabulary size and \(L_y\) signifies the data length.
Existing SemCom systems for multimodal representation often assume that the semantic spaces of different modalities are perfectly aligned, i.e., that one modality can be mapped to another without significant semantic loss.
To formalize the inter-modal semantic relationship, we adopt a theoretical framework grounded in synonymous mapping~\cite{sbcs25/kai}, in which each semantic concept corresponds to a set of syntactic samples sharing the same meaning, termed a synonymous set (detailed in Appendix~\ref{App:Semantic_Theory}). Let $H_s(\cdot,\cdot)$ denote the semantic joint entropy and $I_s(\cdot;\cdot)$ denote the semantic mutual information. Note that the subscript $s$ indicates that these measures are computed over the discrete semantic concepts derived from the raw modalities. 
The quantity $H_s(\mathbf{X}, \mathbf{Y}) - I_s(\mathbf{X}; \mathbf{Y})$ quantifies the total unique semantic information inherent to each modality. Based on this, we define two distinct scenarios. 
\begin{Definition}[Multimodal Semantic Representation Alignment]
    \label{def:alignment}
    A state where no unique semantic information exists in either modality, corresponding to the ideal condition:
    \begin{equation}
        \label{eq.Multimodal_alignment}
        H_s(\mathbf{X},\mathbf{Y}) - I_s(\mathbf{X};\mathbf{Y}) = 0.
    \end{equation}
\end{Definition}
In practice, systems may achieve near-alignment where the unique information is negligibly small. However, strictly perfect alignment rarely holds in real-world scenarios. For instance, in audio--text pairs, acoustic features (e.g., prosodic information) often convey unique semantics absent from the textual transcript.
This leads to the second scenario. 
\begin{Definition}[Multimodal Semantic Representation Gap]
    \label{def:gap} 
    A state where a non-zero amount of unique semantic information exists in at least one modality, defined by the condition:
    \begin{equation}
        \label{eq.Multimodal_gap}
        H_s(\mathbf{X},\mathbf{Y}) - I_s(\mathbf{X};\mathbf{Y}) > 0.
    \end{equation}
\end{Definition}
Since the synonymous sets of each modality are disjoint, the semantic variables are deterministic functions of the sources, i.e., \(\tilde{\mathbf{X}}=g_x(\mathbf{X})\) and \(\tilde{\mathbf{Y}}=g_y(\mathbf{Y})\) (see Appendix~\ref{App:Semantic_Theory}). Hence, \(H_s\) and \(I_s\) are the classical entropy and mutual information evaluated on these coarse-grained variables, e.g., \(I_s(\mathbf{X};\mathbf{Y})=I(\tilde{\mathbf{X}};\tilde{\mathbf{Y}})\le I(\mathbf{X};\mathbf{Y})\) by the data-processing inequality. For digitized sources, the chain rule further gives
\begin{equation}
    \label{eq.entropy_split}
    H(\mathbf{X}) = I_s(\mathbf{X};\mathbf{Y}) + H_s(\mathbf{X}|\mathbf{Y}) + H(\mathbf{X}|\tilde{\mathbf{X}}),
\end{equation}
where \(H(\mathbf{X}|\tilde{\mathbf{X}})\) denotes the intra-concept (syntactic) details that are invisible to semantic measures but required for source recovery.
Under \textbf{Definition}\,\ref{def:gap}, this paper aims to map multimodal data into a unified, compact, and robust semantic representation space for multimodal recovery.
To this end, we propose partitioning the semantic representation space into two distinct components: (\romannumeral1) a shared semantic representation, which targets the information common to both modalities, i.e., \(I_s(\mathbf{X};\mathbf{Y})\); and (\romannumeral2) unique semantic representations, which carry information specific to each modality, i.e., \(H_s(\mathbf{X}|\mathbf{Y})\) for modality \(\mathbf{X}\) and \(H_s(\mathbf{Y}|\mathbf{X})\) for modality \(\mathbf{Y}\), together with the intra-concept details in \eqref{eq.entropy_split}. Because \(g_x\) and \(g_y\) are unknown for natural data, \(H_s\) and \(I_s\) cannot be computed or optimized directly. They specify the target of the decomposition, whereas the optimization in Section\,\ref{Subsec:Problem Formulation} is stated on learned latent variables and linked to them by \textbf{Proposition}\,\ref{prop:link}.

\subsection{Modality-specific Transmitters}
Due to the structural differences between continuous and discrete data, two modality-specific semantic encoders extract semantic representations from each source. 
As illustrated in Fig.\,\ref{fig:framework}, the semantic encoder for continuous input \(\mathbf{x}\) is defined as
\begin{equation}
    \label{eq.img_encoder} 
    (\mathbf{s}_x, \mathbf{u}_x) = \mathcal{F}_x(\mathbf{x}; \boldsymbol{\alpha}),
\end{equation}
where \(\mathcal{F}_x(\cdot; \boldsymbol{\alpha})\) denotes the encoder for continuous data, and \(\boldsymbol{\alpha}\) represents its set of trainable parameters.
Similarly, the semantic encoder for discrete input \(\mathbf{y}\) is given by
\begin{equation}
    \label{eq.text_encoder} 
    (\mathbf{s}_y, \mathbf{u}_y) = \mathcal{F}_y(\mathbf{y}; \boldsymbol{\beta}),
\end{equation}
where \(\mathcal{F}_y(\cdot; \boldsymbol{\beta})\) is the encoder for discrete data and \(\boldsymbol{\beta}\) denotes its trainable parameters.
Specifically, each encoder extracts two types of representations. The shared components, \(\mathbf{s}_x\in\mathbb{R}^{L_s\times D_s}\) and \(\mathbf{s}_y\in\mathbb{R}^{L_s\times D_s}\), capture common semantic information and are constrained to the same dimensions to enable cross-modal alignment and mapping onto a shared knowledge base (Sec.\,\ref{Sec.Knowledge Base}). The unique components, \(\mathbf{u}_x\in\mathbb{R}^{L_{ux}\times D_{u}}\) and \(\mathbf{u}_y\in\mathbb{R}^{L_{uy}\times D_{u}}\), capture modality-specific details~\cite{eccv24/yi}. The lengths \(L_{ux}\) and \(L_{uy}\) may differ, so that each modality can keep its own granularity and structure for reconstruction.
Note that continuous and discrete refer to the source level and only determine the encoder design. At the representation level, for both modalities, shared components are quantized into discrete indices by the shared knowledge base, whereas unique components remain continuous features and are transmitted in analog form.

\subsection{Physical Channel}
\label{Subsec:Physical_Channel}
The extracted representations are transmitted over a noisy wireless channel under imperfect CSI. For brevity, we detail the transmission of \(\mathbf{u}_x\) here. The transmission of \(\mathbf{u}_y\) follows the same procedure. The transmission of \(\mathbf{s}_x\) and \(\mathbf{s}_y\) is deferred to Sec.\,\ref{Sec.Knowledge Base}.
The unique-feature symbols are placed on an orthogonal frequency-division multiplexing (OFDM) resource grid and experience the 3GPP EVA channel~\cite{3gpp.36.101}. Under this power-delay profile, the frequency response varies across subcarriers, and the channel coefficients vary across OFDM symbols according to a Jakes Doppler spectrum with maximum Doppler frequency \(f_D\). On subcarrier \(k\) of OFDM symbol \(t\), the received symbol is
\begin{equation}
    \label{eq.physical_channel}
    y_{k,t} = H_{k,t}\, x_{k,t} + n_{k,t},
\end{equation}
where \(x_{k,t}\in\mathbb{C}\) is the transmitted symbol, \(H_{k,t}\) is the EVA frequency response on that resource element, and \(n_{k,t}\sim\mathcal{CN}(0,\sigma_n^2)\).

Imperfect CSI is modeled at the resource-element level. Let \(\hat{H}_{k,t}\) be the estimated response and \(E_{k,t}\) an independent error term with the same second-order statistics as \(H_{k,t}\). Their relationship is
\begin{equation}
    \label{eq.channel_estimation}
    H_{k,t} = \iota \hat{H}_{k,t} + \sqrt{1-\iota^2}\, E_{k,t},
\end{equation}
where \(\iota\in(0,1]\) is the correlation between the true and estimated responses. The case \(\iota = 1\) corresponds to perfect CSI. For AWGN, \(H_{k,t}=\hat{H}_{k,t}=1\).

Each resource element is equalized by zero-forcing with the estimated response, given by 
\begin{equation}
    \label{eq.channel_factor}
    \hat{x}_{k,t} = y_{k,t} / \hat{H}_{k,t}.
\end{equation}
Transmitted symbols are normalized to unit average power. The reference SNR is defined as
\begin{equation}
    \label{eq.SNR}
    \Gamma = 10\log_{10}\!\left(1 / \sigma_n^2\right).
\end{equation}
With \(\mathbb{E}[|H_{k,t}|^2]=1\), \(\Gamma\) is the average SNR before equalization. Zero forcing leaves residual noise with variance proportional to \(1/|\hat{H}_{k,t}|^2\). The equalized symbols are reassembled into \(\hat{\mathbf{u}}_x\) and passed to the decoder.

\subsection{Unified Receiver}
\label{Subsec:Unified_Receiver}
Since both modalities are mapped into a unified representation space, a single decoder is used at the receiver for source reconstruction.
Let \(\hat{\mathbf{s}}_x\) and \(\hat{\mathbf{u}}_x\) denote the received signals from the continuous modality, corrupted by noise. The decoding process for this modality is defined as
\begin{equation}
    \label{eq.img_decoder}
    \hat{\mathbf{x}} = \mathcal{G} (\hat{\mathbf{s}}_x, \hat{\mathbf{u}}_x; \boldsymbol{\theta}),
\end{equation}
where \(\mathcal{G} (\cdot, \cdot; \boldsymbol{\theta})\) denotes the decoding function, and \(\boldsymbol{\theta}\) represents its associated trainable parameters. 
Similarly, for the discrete modality, the decoded output is given by
\begin{equation}
    \label{eq.tex_decoder}
    \hat{\mathbf{y}} = \mathcal{G} (\hat{\mathbf{s}}_y, \hat{\mathbf{u}}_y; \boldsymbol{\theta}).
\end{equation}
The decoder uses the same parameters \(\boldsymbol{\theta}\) for both modalities. Dimensional differences between \(\hat{\mathbf{u}}_x\) and \(\hat{\mathbf{u}}_y\) are handled by modality-specific projections into a common latent space, and a learnable modality indicator specifies which modality to reconstruct. Although Fig.\,\ref{fig:framework} shows both modalities for clarity, each modality can be transmitted and decoded independently. Cross-modal correlation is used mainly during training to learn the shared codebook, which reduces the channel uses needed for shared semantics at inference. At the receiver, \(\hat{\mathbf{s}}\) and \(\hat{\mathbf{u}}\) are recombined for source reconstruction. Transmitter-side decoupling remains necessary because the shared and unique components are given unequal protection and different channel interfaces.

\begin{figure}[t]
    \centering
    \includegraphics[scale=0.45]{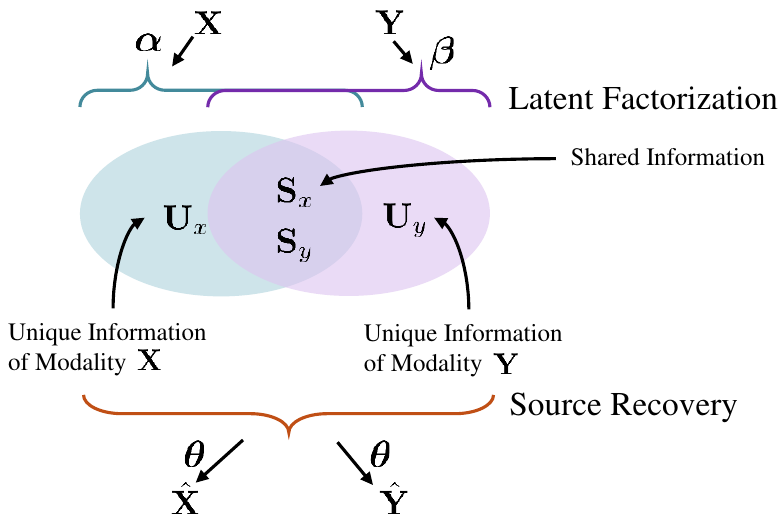}
    \caption{Illustration of multimodal information under the multimodal semantic representation gap.}
    \label{fig:venny}
\end{figure}

\subsection{Problem Formulation}
\label{Subsec:Problem Formulation}
The problem is mathematically formulated by first introducing the necessary notations. Let \(\mathbf{S}_x\) and \(\mathbf{S}_y\) denote the random variables representing the shared information for the continuous data and discrete data, respectively. The unique information associated with the continuous and discrete data is denoted by \(\mathbf{U}_x\) and \(\mathbf{U}_y\). Furthermore, \(I(\cdot;\cdot)\) denotes the classical mutual information. 

As illustrated in Fig.\,\ref{fig:venny}, modalities \(\mathbf{X}\) and \(\mathbf{Y}\) are factorized into shared and unique latent variables by the encoders with parameters \(\boldsymbol{\alpha}\) and \(\boldsymbol{\beta}\), respectively. At the receiver, the sources are recovered by the decoder with parameters \(\boldsymbol{\theta}\).
Therefore, the proposed MSC framework must simultaneously achieve three distinct objectives.
First, the sources must be reconstructed at the receiver. This is pursued by maximizing the mutual information between the original sources \((\mathbf{X}, \mathbf{Y})\) and the reconstructions \((\hat{\mathbf{X}}, \hat{\mathbf{Y}})\), so that the transmitted content is preserved through the channel.
Second, the mutual information between the shared features \(\mathbf{S}_x\) and \(\mathbf{S}_y\) is maximized to align the two modalities in a common representation space.
Finally, the mutual information between the shared features \((\mathbf{S}_x, \mathbf{S}_y)\) and the unique features \((\mathbf{U}_x, \mathbf{U}_y)\) is minimized, so that modality-specific content is kept out of the shared representation~\cite{nips23/paul}.
Therefore, we formulate the overall problem as the following optimization problem.
\begin{Problem}
    \label{Pro.Mutual Information}
    The objective of the proposed multimodal semantic communication system is threefold: (\romannumeral1) source reconstruction, (\romannumeral2) shared semantic information alignment, and (\romannumeral3) unique semantic information decoupling. This is mathematically expressed as:
    \begin{align}
        \mathop{\max}\limits_{\boldsymbol{\alpha}, \boldsymbol{\beta}, \boldsymbol{\theta}} \quad
        & \underbrace{I(\mathbf{X} ; \hat{\mathbf{X}}) + I(\mathbf{Y} ; \hat{\mathbf{Y}})}_{\rm{Reconstruction}} + \underbrace{I(\mathbf{S}_x ; \mathbf{S}_y)}_{\rm{Alignment}} \notag \\
        & \underbrace{ - I(\mathbf{S}_x , \mathbf{S}_y; \mathbf{U}_x , \mathbf{U}_y)}_{\rm{Decoupling}}.
    \end{align}
\end{Problem}
\textbf{Problem}\,\ref{Pro.Mutual Information} is stated on the learned representations, whereas \textbf{Definition}\,\ref{def:gap} is stated on the semantic variables. The two are related as follows.
\begin{Proposition}
    \label{prop:link}
    For the encoders in \eqref{eq.img_encoder} and \eqref{eq.text_encoder}, the following holds:
    \begin{equation}
        \label{eq.leak_shannon}
        I(\mathbf{U}_x;\mathbf{Y}) \le I(\mathbf{S}_x,\mathbf{S}_y;\mathbf{U}_x,\mathbf{U}_y) + I(\mathbf{X};\mathbf{Y}) - I(\mathbf{S}_x;\mathbf{S}_y).
    \end{equation}
    If, in addition, (\romannumeral1) the shared representations \(\mathbf{S}_x\) and \(\mathbf{S}_y\) depend on the sources only through \(\tilde{\mathbf{X}}\) and \(\tilde{\mathbf{Y}}\), and (\romannumeral2) \(\mathbf{X}\to\tilde{\mathbf{X}}\to\tilde{\mathbf{Y}}\) forms a Markov chain, then
    \begin{align}
        \label{eq.align_ceiling}
        & I(\mathbf{S}_x;\mathbf{S}_y) \le I_s(\mathbf{X};\mathbf{Y}), \\
        \label{eq.leak_semantic}
        & I(\mathbf{U}_x;\tilde{\mathbf{Y}}) \le I(\mathbf{S}_x,\mathbf{S}_y;\mathbf{U}_x,\mathbf{U}_y) + I_s(\mathbf{X};\mathbf{Y}) - I(\mathbf{S}_x;\mathbf{S}_y).
    \end{align}
    Symmetric bounds hold for \(\mathbf{U}_y\) under the corresponding conditions with the roles of the two modalities exchanged.
\end{Proposition}
\begin{IEEEproof}
    The proof relies only on the chain rule and the data-processing inequality, using the fact that \(\mathbf{S}_x\) and \(\mathbf{U}_x\) are functions of \(\mathbf{X}\). The details are given in Appendix~\ref{App:Proof_Prop}.
\end{IEEEproof}
By \eqref{eq.align_ceiling}, maximizing the alignment term pushes the shared branch toward the semantic common information \(I_s(\mathbf{X};\mathbf{Y})\) from below. By \eqref{eq.leak_shannon} and \eqref{eq.leak_semantic}, the cross-modal information that leaks into the unique branch is upper-bounded by the decoupling term plus the alignment gap, and \textbf{Problem}\,\ref{Pro.Mutual Information} reduces both. Decoupling alone is therefore insufficient, because any shared content missed by \(\mathbf{S}_x\) must be carried by \(\mathbf{U}_x\) for reconstruction. The reconstruction term, in turn, rules out the trivial solution of an uninformative unique branch. Conditions (\romannumeral1) and (\romannumeral2) are idealizations. They are encouraged, but not enforced, by cross-modal alignment, the many-to-one codebook mapping in Section\,\ref{Sec.Knowledge Base}, and the global teacher features in Section\,\ref{Sec.Knowledge Distillation Aided Multimodal Recovery}, whereas \eqref{eq.leak_shannon} holds without them.
Because these mutual information terms are difficult to compute directly, we derive a tractable training objective from variational information bottleneck principles, with a lower bound for alignment and an upper bound for decoupling, as detailed in Section\,\ref{Sec.System Implementation}. The resulting differentiable loss jointly trains reconstruction, alignment, and decoupling.

\section{Shared Knowledge Base Design}
\label{Sec.Knowledge Base}

\subsection{Shared Knowledge Base Mechanism}
\label{Subsec:Knowledge_Base_Mechanism}
As illustrated in Fig.\,\ref{fig:framework}, the proposed MSC framework uses a shared knowledge base that maps continuous shared representations to discrete codewords, so that only indices need to be transmitted and the communication overhead is reduced. The same codebook also provides a common discrete space for cross-modal alignment and a synchronized dictionary at the transmitter and the receiver. Unlike unimodal knowledge bases for text~\cite{twc24/peng} or images~\cite{tcom25/hongwei}, building one codebook for multimodal data is challenging, mainly because of the multimodal semantic representation gap in \textbf{Definition}\,\ref{def:gap}. For instance, a text-centric codebook may miss image-specific semantics and inject text-side bias, whereas an image-centric codebook may overlook text-specific features.

To this end, we use vector quantization~\cite{iccv23/chuanxia} to quantize the shared components into discrete vectors. Here, ``shared'' has two meanings: (\romannumeral1) semantic interoperability across modalities and (\romannumeral2) synchronization between the transmitter and the receiver. Mathematically, the shared knowledge base is a vector codebook \(\mathcal{K} \triangleq \{\mathbf{k}_j\}_{j=1}^{M}\), where each \(\mathbf{k}_j \in \mathbb{R}^{1 \times D_s}\) is a shared semantic unit and \(M > 1\) is the codebook size. The codebook provides a discrete set of basis elements for cross-modal alignment.

At the transmitter, the shared representations \(\mathbf{s}_x \in \mathbb{R}^{L_s \times D_s}\) and \(\mathbf{s}_y \in \mathbb{R}^{L_s \times D_s}\) are mapped to the discrete codebook. Nearest-neighbor matching under Euclidean distance yields the codebook indices. For brevity, the modality subscript on \(\mathbf{s}\) is omitted in this section, since both modalities use the same codebook and the same mapping. The mapping is
\begin{equation}
\label{eq.Codebook_mapping}
    m_i = \mathop{\arg\min}\limits_{j=1,\dots,M}
    {\left\| \mathbf{s}[i,:] - \mathbf{k}_j \right\|}_2, \quad \text{for } i=1, \ldots, L_s.
\end{equation}
Here, \(m_i\) is the transmitted index for the \(i\)-th row of \(\mathbf{s}\), represented with \(\log_2 M\) bits. The quantized representation is denoted by \(\mathbf{z}\in\mathbb{R}^{L_s\times D_s}\). At the receiver, a correct index retrieves its codeword in \(\mathcal{K}\). An erroneous index is replaced by one fixed learnable embedding, shared by all failed indices and trained with the decoder.

Unlike prior studies~\cite{twc24/peng, tcom25/hongwei} that assume error-free index delivery, the indices here are sent by conventional digital links. Together with analog JSCC of the unique components, the overall system is a hybrid digital-analog SemCom design~\cite{jsac25/huiqiang}: only recurring shared semantics are quantized, while high-dimensional modality-specific residuals remain continuous to limit digital cliffs and preserve fine detail.

\subsection{Shared Knowledge Base Construction}
Constructing the discrete knowledge base \(\mathcal{K}\) involves an \(\arg\min\) operation that is non-differentiable and thus unsuitable for direct end-to-end training. To overcome this, the quantization loss \(\mathcal{L}_{Q}\) is
\begin{equation}
\label{eq:Quantization_Loss}
    \mathcal{L}_{Q} =
    {\| \mathrm{sg}[\mathbf{s}] - {\mathbf{z}} \|}_2^2
    + \lambda_1 {\| \mathbf{s} - \mathrm{sg}[{\mathbf{z}}] \|}_2^2 ,
\end{equation}
where $\mathrm{sg}[\cdot]$ denotes the stop-gradient operation, and $\lambda_1\in\mathbb{R}^+$ is a hyperparameter balancing the two terms.

Stable codebook training is obtained by updating the codebook with exponential moving average (EMA)~\cite{iccv23/chuanxia}. We also use \(\ell_2\)-normalized distance for nearest-code search and reduce the codebook embedding dimension to \(32\)~\cite{iccv23/chuanxia}. These low-dimensional embeddings are then projected back to a higher-dimensional space before decoding. Accordingly, the mapping in \eqref{eq.Codebook_mapping} is replaced by
\begin{equation}
    \label{eq.Codebook_mapping_revised}
    \begin{aligned}
        m_i = \mathop{\arg\min}\limits_{j=1,\dots,M}
        &\bigl\| \ell_2\bigl(f(\mathbf{s}[i,:])\bigr) - \ell_2\bigl(f(\mathbf{k}^{j})\bigr) \bigr\|_2,\\
        &\qquad i=1,\ldots,L_s.
    \end{aligned}
\end{equation}
A dynamic revival strategy is further used to mitigate underutilized codebook entries (``dead codes'') under EMA updates. A code is treated as dead if its EMA-maintained cluster size falls below a threshold \(\tau\). Instead of reinitializing dead codes with random batch samples, we perturb active codes from high-utilization clusters: a selected active code is perturbed by additive Gaussian noise \(\epsilon \sim \mathcal{N}(0, 0.01)\) and then \(\ell_2\)-normalized. This keeps the revived entries near well-learned regions of the codebook and improves utilization relative to random reinitialization.

After training, the codebook is frozen and deployed at both the transmitter and the receiver together with the encoder and decoder, so that both ends hold an identical copy. During transmission, only the indices are exchanged, and no codebook synchronization is required. Because the codebook is jointly optimized with the encoder and decoder, it is kept as part of the model. When an update is needed, e.g., under a data distribution shift, the codebook is fine-tuned together with the codec and redeployed to both ends as a new model version. Mismatched codebooks would map received indices to incorrect codewords, so the codebook, encoder, and decoder are always updated jointly. Dynamic codebook updating and synchronization after deployment are left for future work.

\section{Knowledge Distillation-Aided Multimodal Recovery}
\label{Sec.Knowledge Distillation Aided Multimodal Recovery}
In SemCom, multimodal understanding and recovery are often treated as separate tasks. Recent work~\cite{nips24/luting} suggests that understanding can benefit recovery, and empirical SemCom studies show that coupling the two, in series or in parallel at the receiver, can improve overall performance. Such pipelines are usually limited by supervised training: large-scale labels are scarce, and discrete labels often cannot capture complex semantics.

To overcome this challenge, we propose employing knowledge distillation, a technique originally developed to transfer knowledge from computationally intensive teacher models to more lightweight student models. In our framework, knowledge distillation aligns intermediate representations through feature reconstruction, so that the student encoder learns semantically informative features from the teacher model. Note that the semantic encoder in our framework serves as the student model. 

\begin{figure}
    \centering
    \includegraphics[scale=0.5]{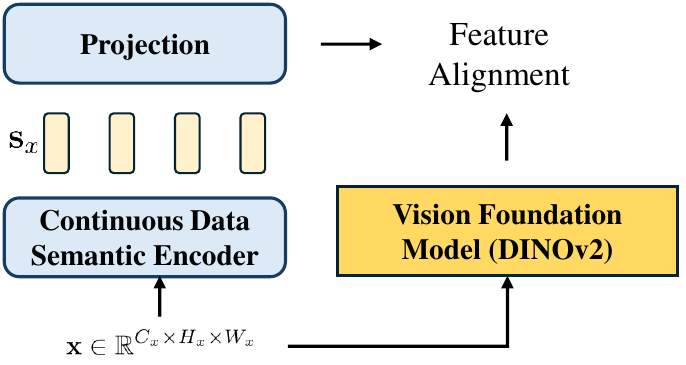}
    \caption{A demonstration of knowledge distillation, in which a vision foundation model is used to align the features output by the semantic encoder. A projection module with two linear layers is used to align the dimensions.}
    \label{fig:alignment}
\end{figure}

Specifically, the pretrained model DINOv2~\cite{tmlr24/maxime}, denoted as \(\mathcal{T}(\cdot;\boldsymbol{\gamma}_t)\), serves as the teacher. Pretrained with self-supervision on large-scale data, DINOv2 extracts rich semantics from an input image \(\mathbf{x}\).
We adopt the frozen ViT-S/14 variant, which takes the same input image as the MSC encoder. Only the global information is used, namely the final-layer CLS feature, as the teacher signal for the shared image components.
The feature extraction by the teacher model is formulated as
\begin{equation}
    \mathbf{z}_x^{t} = \mathcal{T}(\mathbf{x};\boldsymbol{\gamma}_t),
\end{equation}
where $\mathbf{z}_x^{t}\in\mathbb{R}^{1\times D_t}$ represents the semantic features extracted by the teacher model from the input $\mathbf{x}$, and $D_t=384$ is the teacher feature dimension.

Meanwhile, the quantized shared representation \(\mathbf{z}_x\in\mathbb{R}^{L_s\times D_s}\) is flattened to \(\mathrm{vec}(\mathbf{z}_x)\) and fed to a projection module \(\mathcal{P}(\cdot;\boldsymbol{\theta}_p)\) with parameters \(\boldsymbol{\theta}_p\), implemented as two linear layers. This module maps the codebook features to the teacher feature space, given by
\begin{equation}
    \mathbf{z}_x^{s} = \mathcal{P}(\mathrm{vec}(\mathbf{z}_x);\boldsymbol{\theta}_p),
\end{equation}
where \(\mathbf{z}_x^{s}\in\mathbb{R}^{1\times D_t}\) is the projected student feature.

The distillation loss \(\mathcal{L}_{\text{KD}}\) encourages the student feature \(\mathbf{z}_x^{s}\) to align with the teacher feature \(\mathbf{z}_x^{t}\). Cosine distance is used for directional alignment. An MSE term is further added between the quantized shared features \(\mathbf{z}_x\) and \(\mathbf{z}_y\), so that the teacher's guidance on the image branch also regularizes the text branch. This loss is defined as
\begin{equation}
    \begin{aligned}
    \mathcal{L}_{\text{KD}} = \frac{1}{N}\sum_{i=1}^{N} \Bigg[ & \left(1 - \frac{\mathbf{z}_x^{t,i} {\mathbf{z}_x^{s,i}}^{T}} {{\|\mathbf{z}_x^{t,i}\|}_2{\|\mathbf{z}_x^{s,i}\|}_2}\right) \\
    & + \lambda_2\|\mathrm{sg}[\mathbf{z}_x^{i}] - \mathbf{z}_y^{i} \|_2^2 \Bigg],
    \end{aligned}
\end{equation}
where \(\lambda_2\in\mathbb{R}^+\) is a hyperparameter, the superscript \(i\) indexes mini-batch samples, and \(\mathrm{sg}[\cdot]\) is the stop-gradient in \eqref{eq:Quantization_Loss}. The stop-gradient lets the teacher-guided image codewords act as anchors for the text codewords. Both \(\mathcal{T}\) and \(\mathcal{P}\) are used only in training and are discarded at inference. Let \(\mathbf{Z}_x\) and \(\mathbf{Z}_y\) denote the random variables of the quantized shared representations. When \(\mathbf{z}_x=\mathbf{z}_y\), the shared codeword is a common function of both modalities, so that \(H(\mathbf{Z}_x|\mathbf{Y})=H(\mathbf{Z}_y|\mathbf{X})=0\). The MSE term therefore also discourages modality-specific content in the shared branch, complementing the leakage bounds in \textbf{Proposition}\,\ref{prop:link} for the unique branch.

\begin{figure}[t]
    \centering
    \includegraphics[scale=0.50]{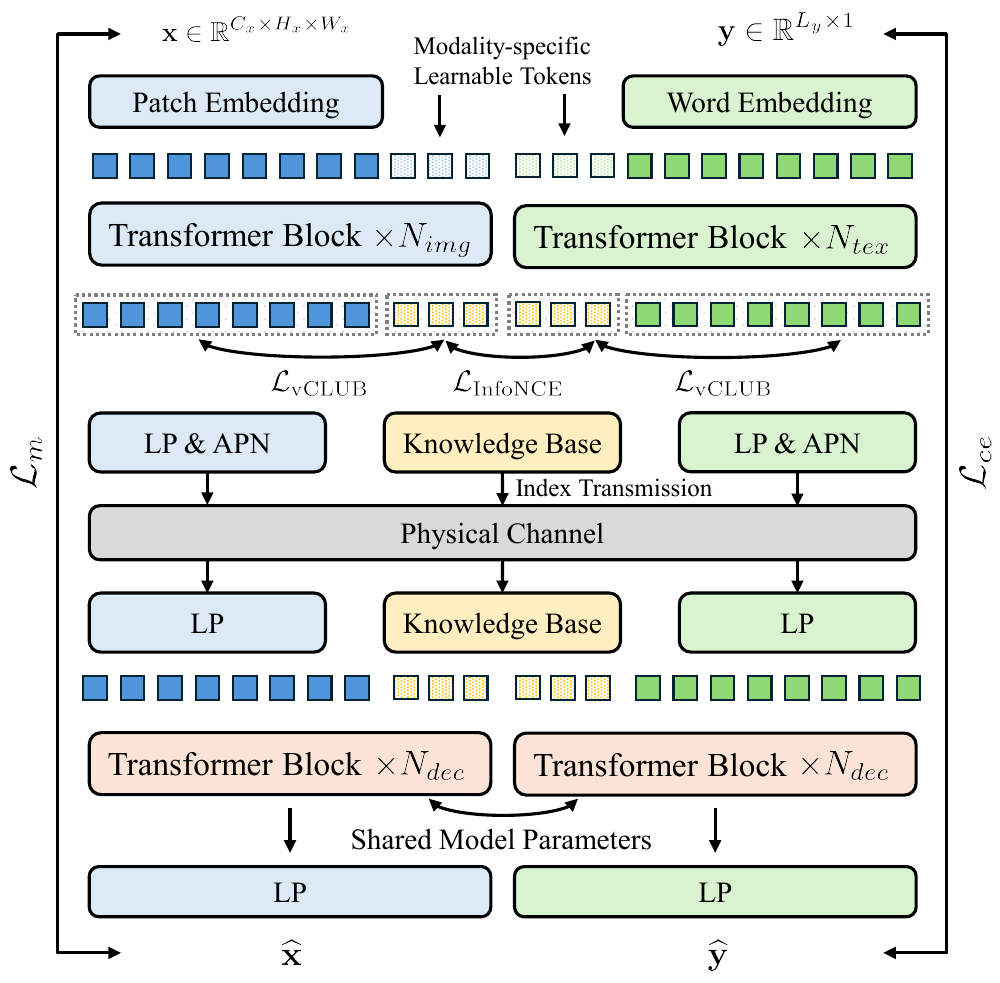}
    \caption{The proposed network structure of the MSC system. ``LP'' and ``APN'' stand for linear projection and average power normalization, respectively.}
    \label{Fig.network}
\end{figure}

\section{Proposed MSC System Implementation}
\label{Sec.System Implementation}

\subsection{Network Structure}
\subsubsection{Transmitter}
The proposed MSC system builds a unified semantic representation space for heterogeneous inputs. Each modality is first converted into a sequence of embeddings~\cite{wc25/li}.
The network is shown in Fig.~\ref{Fig.network}. For images, a patch embedding module splits the input into fixed-size patches, flattens them, and applies a linear projection to form image embeddings. For text, each word is mapped to a dense vector by a word embedding layer. Positional encodings are then added to both sequences.

To separate shared semantics from the inputs, \(N_{c}\) learnable embeddings are concatenated to the input sequence before the modality-specific Transformer encoders. The image and text Transformers have \(N_{img}\) and \(N_{tex}\) blocks, respectively. With multi-head self-attention, the encoders extract modality-specific features \(\mathbf{u}\) while aggregating shared semantics into \(\mathbf{s}\). 
Layer normalization is applied before each sub-layer, and positional encodings are added to the query and key matrices in attention.

\subsubsection{Channel}
After encoding, the outputs are split into shared features \((\mathbf{z}_x / \mathbf{z}_y)\) and modality-specific features \((\mathbf{u}_x / \mathbf{u}_y)\). The discrete indices of the shared features are converted to bit sequences, protected by rate-\(1/2\) low-density parity-check (LDPC) codes, and modulated with quadrature phase-shift keying (QPSK).
The continuous unique features \((\mathbf{u}_x / \mathbf{u}_y)\) are sent by analog JSCC. A linear projection first compresses them to a lower-dimensional latent space. The resulting real vectors are treated as the real and imaginary parts of complex baseband symbols, and a normalization layer enforces the average power constraint. These complex symbols are then mapped onto the OFDM resource grid. Each symbol occupies one resource element and is scaled by the EVA coefficient of that element, after which the receiver applies \eqref{eq.channel_factor}. For an image, each embedding uses one subcarrier and each complex dimension uses one OFDM symbol. Valid text symbols are written in subcarrier-major order, and padding symbols are not transmitted. At the receiver, a symmetric linear projection maps the received complex symbols back to the original feature dimension for decoding.

\subsubsection{Receiver}
At the receiver, a shared Transformer decoder with \(N_{dec}\) layers reconstructs both modalities from the noisy received features. The Transformer backbone is shared, while modality-specific linear projections and a learnable modality indicator make the decoder modality-aware, so sequences with different unique lengths (\(L_{ux}\) versus \(L_{uy}\)) can share parameters \(\boldsymbol{\theta}\). The indicator is prepended to the received sequence and guides projection back to each modality's feature space. A final linear layer maps the decoder output to the data space (pixels for images, vocabulary logits for text).

\subsection{Objective Functions}
To address \textbf{Problem}\,\ref{Pro.Mutual Information}, the mutual information objectives are turned into loss functions that update \(\boldsymbol{\alpha}\), \(\boldsymbol{\beta}\), \(\boldsymbol{\theta}\), and the knowledge base \(\mathcal{K}\) by gradient descent. We consider a dataset of \(N\) multimodal pairs \(\{(\mathbf{x}^i, \mathbf{y}^i)\}_{i=1}^{N}\), where each pair has partially overlapping semantics.

\subsubsection{Reconstruction}
To ensure effective source data recovery, the transmitter and receiver are jointly optimized. The primary objective is to maximize the mutual information between the original inputs and their reconstructions, i.e., \(I(\mathbf{X}; \hat{\mathbf{X}})\) and \(I(\mathbf{Y}; \hat{\mathbf{Y}})\). 
For images, pixel-level distortion is used instead of bit error rate (BER). If the reconstruction error is modeled as Gaussian, maximizing \(I(\mathbf{X}; \hat{\mathbf{X}})\) is equivalent to minimizing the mean squared error (MSE). Under a Laplacian model, it is equivalent to minimizing the mean absolute error (MAE). We therefore use a weighted hybrid image loss, formulated as
\begin{equation}
\label{eq.m_Loss}
    \mathcal{L}_{\text{m}} = \frac{1}{N} \sum_{i=1}^{N} \left[ { \left\| \mathbf{x}^i - \hat{\mathbf{x}}^i \right\|_1 }
    + \lambda_3 {\left\| \mathbf{x}^i - \hat{\mathbf{x}}^i \right\|_2^2 } \right],
\end{equation}
where $N$ is the number of samples, $\mathbf{x}^i$ and $\hat{\mathbf{x}}^i$ are the original and reconstructed image data for the $i$-th sample, respectively, and $\lambda_3 \in \mathbb{R}^+$ is a weighting factor that balances the MAE ($\|\cdot\|_1$) and MSE ($\|\cdot\|_2^2$) components.
Text recovery is treated as classification with cross-entropy (CE) loss, given by
\begin{equation}
    \mathcal{L}_{\text{ce}} = \frac{1}{N L_y} \sum_{i=1}^{N} \sum_{l_y=0}^{L_y -1} -\log {\hat{\mathbf{y}}^i \left[l_y, \mathbf{y}^i [l_y]\right]},
\end{equation}
where \(L_y\) is the text length, \(\mathbf{y}^i[l_y]\) is the true word index at position \(l_y\), and \(\hat{\mathbf{y}}^i[l_y, \mathbf{y}^i[l_y]]\) is the predicted probability of that index.

\subsubsection{Alignment}
A critical objective in the MSC system is to align the shared representation spaces derived from different modalities. This alignment is achieved by maximizing the mutual information \(I(\mathbf{S}_x; \mathbf{S}_y)\) between the semantic representations \(\mathbf{S}_x\) (from modality \(x\)) and \(\mathbf{S}_y\) (from modality \(y\)). 
As in~\cite{icml19/ben}, this mutual information admits the lower bound, expressed as
\begin{equation}
    I(\mathbf{S}_x; \mathbf{S}_y) \geq \mathbb{E} \left[
    \frac{1}{N}\sum_{i=1}^{N} \log \frac{p(\mathbf{s}_y^i | \mathbf{s}_x^i)}{\frac{1}{N}\sum_{j=1}^{N} p(\mathbf{s}_y^i | \mathbf{s}_x^j)} \right],
\end{equation}
where \(p(\mathbf{s}_y^i | \mathbf{s}_x^i)\) represents the conditional likelihood of \(\mathbf{s}_y^i\) given its paired \(\mathbf{s}_x^i\). The expectation is over $N$ samples drawn i.i.d. from the joint data distribution $p(\mathbf{S}_x, \mathbf{S}_y)$. 
Maximizing this bound is equivalent to minimizing the InfoNCE loss~\cite{cvpr20/kaiming}, defined as
\begin{equation}
\label{eq:infoNCE}
    \mathcal{L}_{\text{InfoNCE}} = -\frac{1}{N}\sum_{i=1}^{N} \log \frac{\exp\left( {\mathbf{s}_x^i}^T \mathbf{s}_y^i / \tau \right)}{\sum_{j=1}^{N} \exp\left( {\mathbf{s}_x^i}^T \mathbf{s}_y^j / \tau \right)},
\end{equation}
where \(\tau\in\mathbb{R}^+\) is a temperature hyperparameter. Minimizing \(\mathcal{L}_{\text{InfoNCE}}\) pulls paired features \((\mathbf{s}_x^i,\mathbf{s}_y^i)\) together and pushes unmatched pairs apart, which aligns the shared representations across modalities.

\subsubsection{Decoupling}
To disentangle shared and modality-specific features, \(I(\mathbf{S}_x, \mathbf{S}_y; \mathbf{U}_x, \mathbf{U}_y)\) is minimized. Direct estimation is intractable without closed-form densities. Following~\cite{icml20/pengyu}, a variational upper bound is formed with an approximator \(q_{\boldsymbol{\varpi}}(\mathbf{s}_x, \mathbf{s}_y | \mathbf{u}_x, \mathbf{u}_y)\) parameterized by weights \(\boldsymbol{\varpi}\), as in \eqref{eq.MI Upper}.
The bound holds when \(q_{\boldsymbol{\varpi}}(\mathbf{s}_x,\mathbf{s}_y|\mathbf{u}_x,\mathbf{u}_y)\,p(\mathbf{u}_x,\mathbf{u}_y)\) is closer to the joint distribution \(p(\mathbf{s}_x,\mathbf{s}_y,\mathbf{u}_x,\mathbf{u}_y)\) than the product of marginals \(p(\mathbf{s}_x,\mathbf{s}_y)\,p(\mathbf{u}_x,\mathbf{u}_y)\) in the sense of~\cite{icml20/pengyu}. To this end, the variational log-likelihood is maximized, written as
\setcounter{MYtempeqncnt}{\value{equation}}
\addtocounter{equation}{1}
\begin{equation}
\label{eq.varpi}
   \mathcal{L}_{\boldsymbol{\varpi}} = \frac{1}{N} \sum_{i=1}^N \log q_{\boldsymbol{\varpi}}(\mathbf{s}_x^i, \mathbf{s}_y^i | \mathbf{u}_x^i, \mathbf{u}_y^i).
\end{equation}
The network \(q_{\boldsymbol{\varpi}}\) uses Transformer blocks with self-attention to model dependence between \((\mathbf{S}_x, \mathbf{S}_y)\) and \((\mathbf{U}_x, \mathbf{U}_y)\). After this approximator is trained, the variational CLUB (vCLUB) loss in \eqref{eq.vCLUB Loss} is computed. Minimizing \(\mathcal{L}_{\text{vCLUB}}\) reduces mutual information between shared and unique components and encourages decoupling.
Since \(\log N-\mathcal{L}_{\text{InfoNCE}}\le I(\mathbf{S}_x;\mathbf{S}_y)\) and \(\mathcal{L}_{\text{vCLUB}}\) upper-bounds \(I(\mathbf{S}_x,\mathbf{S}_y;\mathbf{U}_x,\mathbf{U}_y)\) when \(q_{\boldsymbol{\varpi}}\) is accurate, minimizing both losses reduces the two controllable terms on the right-hand sides of \eqref{eq.leak_shannon} and \eqref{eq.leak_semantic}. Note that these estimators bound the classical mutual information of the learned representations. They do not estimate \(H_s\) or \(I_s\), and the guarantees in \textbf{Proposition}\,\ref{prop:link} therefore hold up to the estimation gaps of InfoNCE and vCLUB.

\begin{figure*}[t]
    \centering
    \setcounter{equation}{\value{MYtempeqncnt}}
    \begin{equation}
        \label{eq.MI Upper}
        I(\mathbf{S}_x , \mathbf{S}_y;\mathbf{U}_x , \mathbf{U}_y) 
        \leq \mathbb{E}_{p(\mathbf{s}_x, \mathbf{s}_y, \mathbf{u}_x , \mathbf{u}_y)} [\log {q_{\boldsymbol{\varpi}}(\mathbf{s}_x , \mathbf{s}_y | \mathbf{u}_x , \mathbf{u}_y)}] 
        - \mathbb{E}_{p(\mathbf{s}_x, \mathbf{s}_y)} \mathbb{E}_{p(\mathbf{u}_x , \mathbf{u}_y)} [\log {q_{\boldsymbol{\varpi}}(\mathbf{s}_x , \mathbf{s}_y | \mathbf{u}_x , \mathbf{u}_y)}]
    \end{equation}
    \addtocounter{equation}{1}
    \begin{equation}
    \label{eq.vCLUB Loss}
        \mathcal{L}_{\text{vCLUB}} = \frac{1}{N^2} \sum_{i=1}^N \sum_{j=1}^N \left[ \log q_{\boldsymbol{\varpi}} \left(\mathbf{s}_x^i , \mathbf{s}_y^i | \mathbf{u}_x^i , \mathbf{u}_y^i\right) - \log q_{\boldsymbol{\varpi}} \left(\mathbf{s}_x^j , \mathbf{s}_y^j | \mathbf{u}_x^i , \mathbf{u}_y^i \right) \right] 
    \end{equation}
    \hrulefill
    \vspace{-0.2cm}
\end{figure*}

The overall training loss is the weighted sum, computed as
\begin{equation}
\label{eq.Total Loss}
    \begin{aligned}
    \mathcal{L}_{\text{tot}} = &\lambda_4 \mathcal{L}_{\text{m}} + \lambda_5 \mathcal{L}_{\text{ce}} + \lambda_6 \mathcal{L}_{\text{InfoNCE}} \\ &+ \lambda_7 \mathcal{L}_{\text{vCLUB}} + \lambda_8 \mathcal{L}_{Q} + \lambda_9 \mathcal{L}_{\text{KD}},
    \end{aligned}
\end{equation}
where \(\lambda_i\in\mathbb{R}^+\) for \(i\in\{4,\ldots,9\}\) are hyperparameters.

\subsection{Training Algorithm}
Algorithm~\ref{alg.train} uses a two-stage update in each iteration. Stage~1 updates the variational mutual-information estimator. Stage~2 updates the encoders, the unified decoder, the knowledge distillation projection module, and the shared knowledge base under the total loss. The shared indices are sent digitally and the unique features are sent by analog JSCC, matching the hybrid transmission used at inference.

\begin{algorithm}[t]
    \caption{Training Procedure for the Proposed MSC System.}
    \begin{algorithmic}[1]
    \REQUIRE Training dataset \(\mathfrak{R}\); initial parameters \(\boldsymbol{\alpha}\), \(\boldsymbol{\beta}\), \(\boldsymbol{\theta}\), \(\boldsymbol{\theta}_p\), \(\boldsymbol{\varpi}\); initial shared knowledge base \(\mathcal{K}\); frozen teacher \(\mathcal{T}(\cdot;\boldsymbol{\gamma}_t)\).
    \ENSURE Optimized parameters \(\boldsymbol{\alpha}^{\ast}\), \(\boldsymbol{\beta}^{\ast}\), \(\boldsymbol{\theta}^{\ast}\), \(\boldsymbol{\theta}_p^{\ast}\), and knowledge base \(\mathcal{K}^{\ast}\).
        \FOR{each training iteration}
            \STATE Sample a mini-batch \(\{(\mathbf{x}^i, \mathbf{y}^i)\}_{i=1}^{N}\) from \(\mathfrak{R}\).
            \STATE \textit{// Stage 1: Update the variational mutual information estimator}
            \FOR{$N_{mi}$ steps}
                \STATE Compute \(\mathcal{L}_{\boldsymbol{\varpi}}\) using \eqref{eq.varpi}.
                \STATE Update \(\boldsymbol{\varpi} \leftarrow \boldsymbol{\varpi} + \eta_1 \nabla_{\boldsymbol{\varpi}} \mathcal{L}_{\boldsymbol{\varpi}}\).
            \ENDFOR
            \STATE \textit{// Stage 2: Update the MSC model}
            \STATE \((\{\mathbf{s}_x^i, \mathbf{u}_x^i\}_{i=1}^{N}) \leftarrow\) encode \(\{\mathbf{x}^i\}_{i=1}^{N}\) by \eqref{eq.img_encoder}.
            \STATE \((\{\mathbf{s}_y^i, \mathbf{u}_y^i\}_{i=1}^{N}) \leftarrow\) encode \(\{\mathbf{y}^i\}_{i=1}^{N}\) by \eqref{eq.text_encoder}.
            \STATE Quantize shared features to indices and codewords \(\{\mathbf{z}_x^i, \mathbf{z}_y^i\}_{i=1}^{N}\) via \eqref{eq.Codebook_mapping_revised}.
            \STATE Transmit shared indices with LDPC-coded QPSK; map recovered indices to codebook vectors \(\{\hat{\mathbf{s}}_x^i, \hat{\mathbf{s}}_y^i\}_{i=1}^{N}\).
            \STATE Transmit unique features via \eqref{eq.physical_channel} and equalize via \eqref{eq.channel_factor} to obtain \(\{\hat{\mathbf{u}}_x^i, \hat{\mathbf{u}}_y^i\}_{i=1}^{N}\).
            \STATE Reconstruct \(\{\hat{\mathbf{x}}^i, \hat{\mathbf{y}}^i\}_{i=1}^{N}\) from \((\hat{\mathbf{s}},\hat{\mathbf{u}})\) via \eqref{eq.img_decoder} and \eqref{eq.tex_decoder}.
            \STATE Compute \(\mathcal{L}_{\text{tot}}\) using \eqref{eq.Total Loss} (including \(\mathcal{L}_{Q}\) and \(\mathcal{L}_{\text{KD}}\)).
            \STATE Update
            \((\boldsymbol{\alpha}, \boldsymbol{\beta}, \boldsymbol{\theta}, \boldsymbol{\theta}_p)
            \leftarrow
            (\boldsymbol{\alpha}, \boldsymbol{\beta}, \boldsymbol{\theta}, \boldsymbol{\theta}_p)
            - \eta_2 \nabla_{(\boldsymbol{\alpha}, \boldsymbol{\beta}, \boldsymbol{\theta}, \boldsymbol{\theta}_p)}\mathcal{L}_{\text{tot}}\).
            \STATE Update \(\mathcal{K}\) by EMA.
        \ENDFOR
    \end{algorithmic}
    \label{alg.train}
\end{algorithm}

\section{Simulations and Results}
\label{Sec.Simulations and Results}

\subsection{Simulation Setup}
\subsubsection{Dataset}
All simulations use the COCO Caption dataset~\cite{arxiv15/xinlei}, which contains \(123{,}287\) images, each paired with five human-written captions. The captions describe visual content and thus induce partial semantic overlap between image and text. During training, images are randomly cropped to \(224\times 224\).

\subsubsection{Network Configuration}
Each image is split into a \(16\times 16\) grid of non-overlapping \(14\times 14\) patches. To aid shared representation learning, \(N_c=4\) learnable embeddings are appended to each modality sequence. The image encoder has \(4\) Transformer blocks of dimension \(384\), the text encoder has \(2\) blocks of dimension \(384\), and the unified decoder has \(4\) blocks of dimension \(384\). A linear layer at the transmitter projects image and text unique features to dimensions \(32\) and \(8\), respectively. The knowledge base has \(M=1024\) vectors of dimension \(D_s=32\). Shared indices are sent with LDPC-coded QPSK, forming a hybrid digital-analog link together with the analog unique features.

\subsubsection{Training Configuration}
Models are trained separately for AWGN and 3GPP EVA channels. For EVA, we set \(f_D=70\)\,Hz, \(256\) subcarriers, sampling rate \(15.36\)\,MHz (\(\Delta f=60\)\,kHz), and \(\iota=0.92\). Training SNR is drawn uniformly from \(0\) to \(12\)\,dB.
The weighting hyperparameters for the constituent loss functions, $[\lambda_1, \dots, \lambda_9]$, are set to $[0.25, 0.5, 10.0, 1.0, 0.5, 0.3, 0.1, 10.0, 0.5]$. The mutual information estimator and the proposed model are optimized using separate AdamW optimizers with learning rates $\eta_1 = 5 \times 10^{-4}$, $\eta_2 = 1 \times 10^{-5}$, and a mini-batch size of $400$. DINOv2 is used only as a frozen teacher during training; it is not required at inference.
The quantitative performance results are averaged over three independent training runs with different random seeds.

\begin{figure*}[t]
    \centering
    \captionsetup[subfigure]{labelformat=empty}
    \subfloat[\label{fig.awgn snr PSNR}]{
	\includegraphics[scale =0.36]{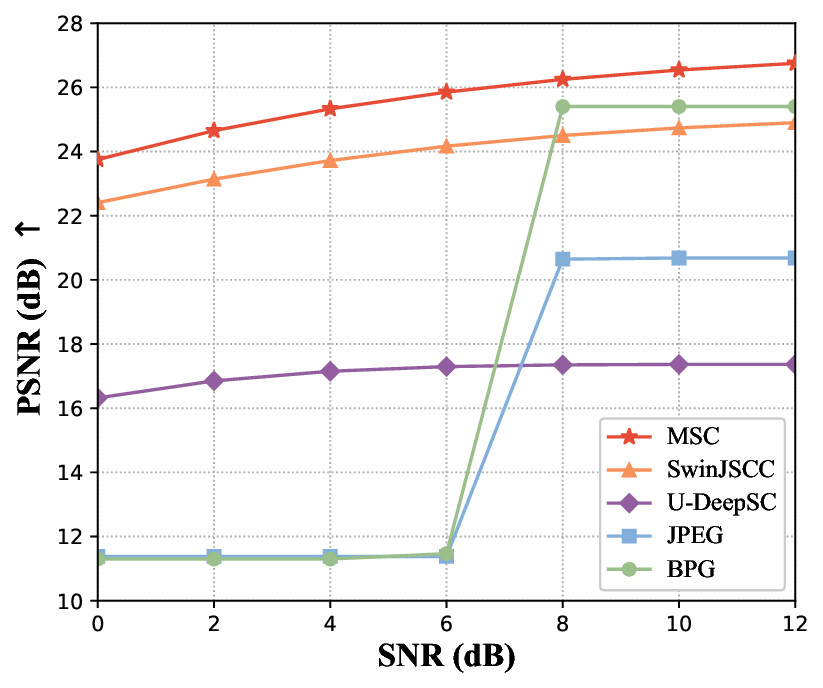}} 
    % \hfill 
    \subfloat[\label{fig.awgn snr SSIM}]{
        \includegraphics[scale =0.36]{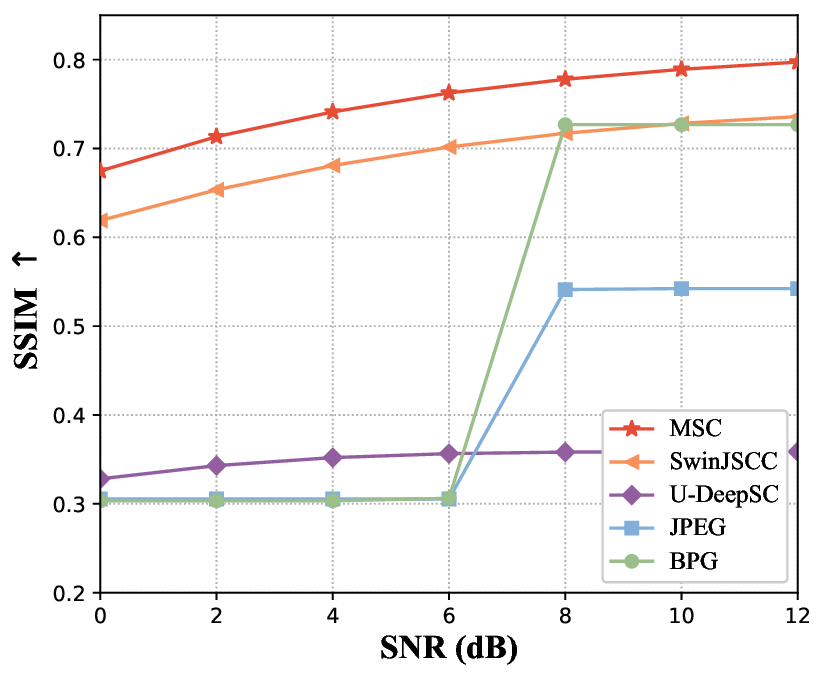}}
    % \hfill 
    \subfloat[\label{fig.awgn snr LPIPS}]{
        \includegraphics[scale =0.36]{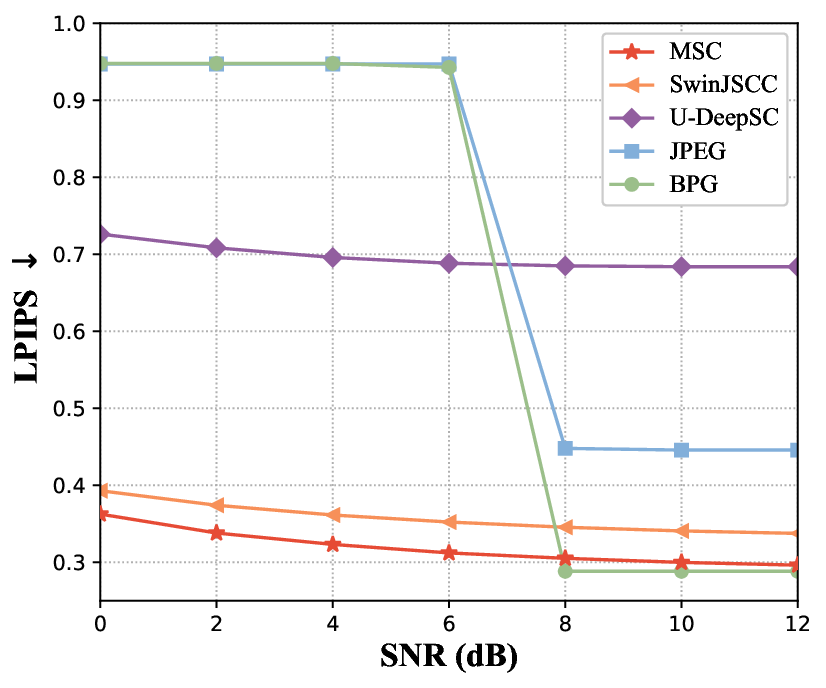}}
    \vspace{-0.5cm}
    \caption{The performance of image transmission versus SNR over an AWGN channel, evaluated in terms of (a) PSNR, (b) SSIM, and (c) LPIPS.}
    \label{fig:awgn snr img}
    \vspace{-0.4cm}
\end{figure*}

\begin{figure*}[t]
    \centering
    \captionsetup[subfigure]{labelformat=empty}
    \subfloat[\label{fig.awgn snr WER}]{
	\includegraphics[scale =0.36]{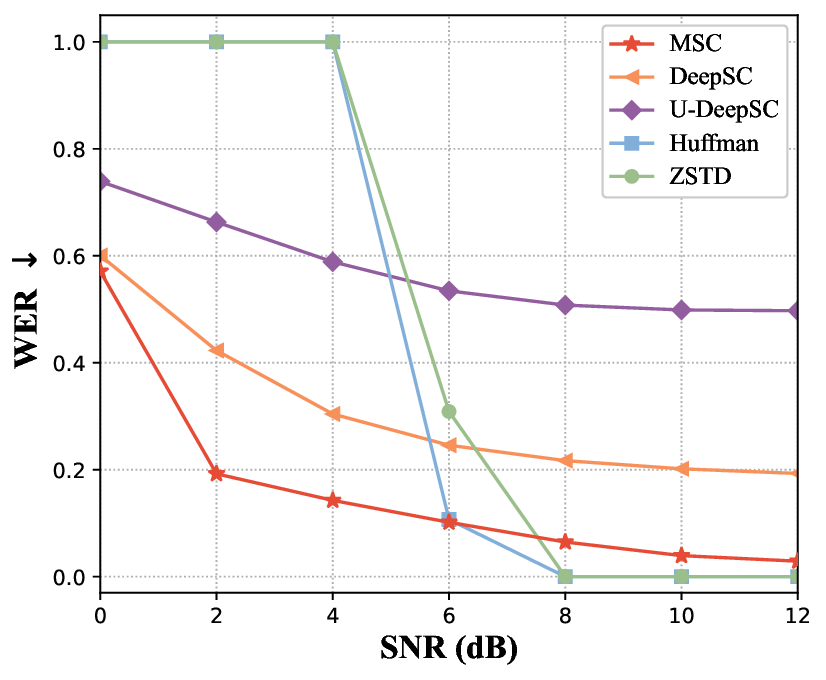}} 
    % \hfill 
    \subfloat[\label{fig.awgn snr BLEU}]{
        \includegraphics[scale =0.36]{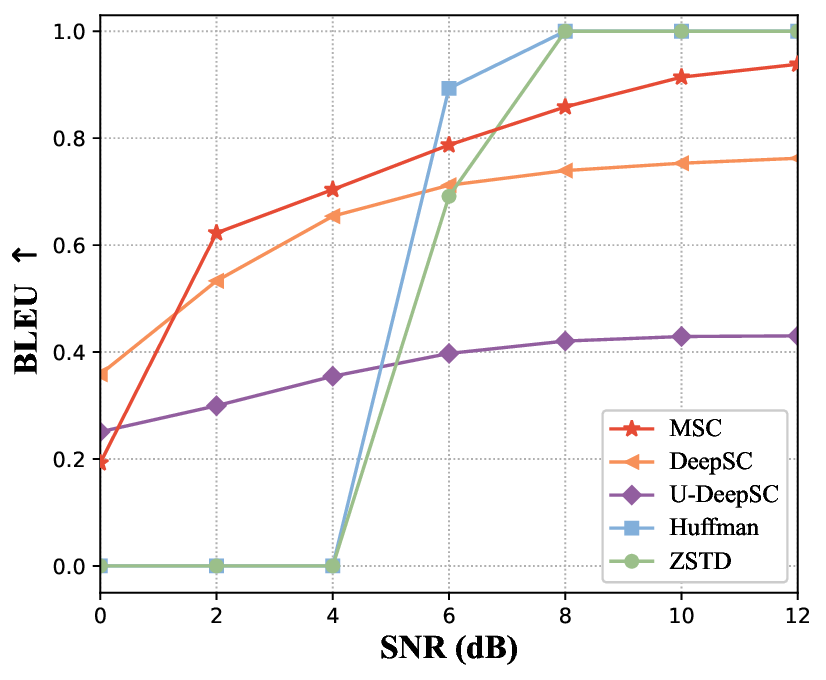}}
    % \hfill 
    \subfloat[\label{fig.awgn snr SS}]{
        \includegraphics[scale =0.36]{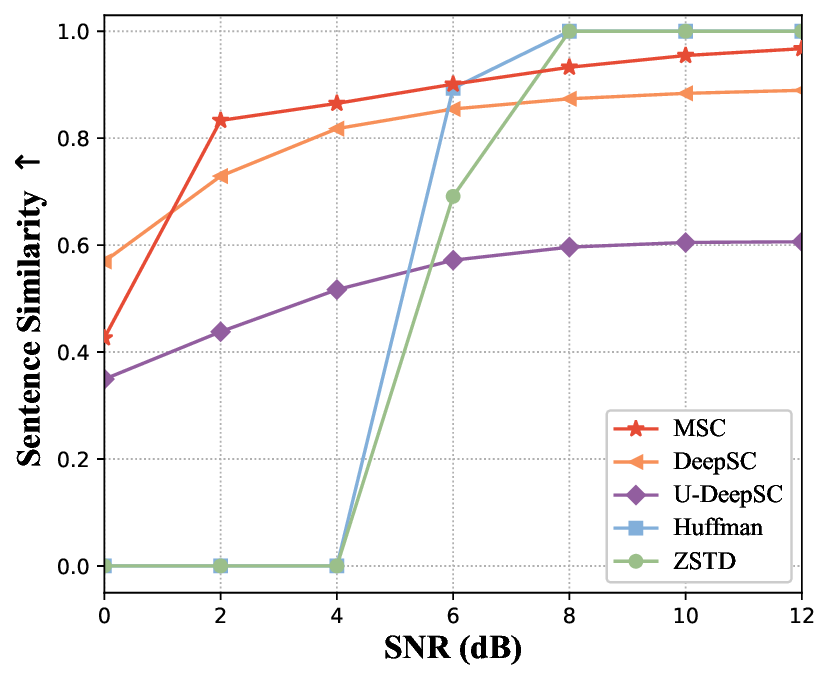}}
    \vspace{-0.5cm}
    \caption{The performance of text transmission versus SNR over an AWGN channel, evaluated in terms of (a) WER, (b) BLEU, and (c) sentence similarity.}
    \label{fig:awgn snr tex}
    \vspace{-0.2cm}
\end{figure*}

\subsubsection{Benchmark Schemes}
We compare the proposed scheme with conventional and learning-based baselines.
\begin{itemize} 
    \item {Separate Source-Channel Coding (SSCC):} Images use Joint Photographic Experts Group (JPEG) or Better Portable Graphics (BPG) source coding. Text uses Huffman or Zstandard (ZSTD). Channel coding uses quasi-cyclic LDPC codes~\cite{3gpp.38.212} with rate \(1/2\) and \(30\) decoding iterations. Block sizes are \(4096\) for images and \(1024\) for text, with 16-QAM. Comparison fairness is defined by matching the total number of complex channel uses.
    \item {Joint Source-Channel Coding (JSCC):} We include SwinJSCC~\cite{tccn25/ke} for images and DeepSC~\cite{tsp21/huiqiang} for text. U-DeepSC~\cite{tcom24/guangyi} is included as a multimodal baseline. We adapt it to image--text reconstruction on COCO, with separate codebooks and uncoded 16-QAM for the quantized indices. All models are trained on the same dataset and under the same channel conditions as MSC.
\end{itemize}

\subsubsection{Performance Metrics}
Unlike bit-level metrics such as BER, we evaluate semantic reconstruction quality as follows.
\begin{itemize}
\item {Peak Signal-to-Noise Ratio (PSNR):} Measures reconstruction fidelity based on pixel-wise mean squared error.
\item {Structural Similarity Index Measure (SSIM):} Assesses perceptual quality by evaluating luminance, contrast, and structural information.
\item {Learned Perceptual Image Patch Similarity (LPIPS)~\cite{cvpr18/richard}:} A deep feature-based metric that measures perceptual similarity, correlating well with human judgment.
\end{itemize}

For text transmission, performance is assessed using the following metrics.
\begin{itemize}
\item {Word Error Rate (WER):} Measures the number of substitutions, deletions, and insertions required to transform the received sentence into the reference sentence.
\item {Bilingual Evaluation Understudy (BLEU)~\cite{acl02/kishore}:} Evaluates word-level accuracy by comparing n-gram overlap. We report BLEU, which is the geometric mean of precisions for n-grams of sizes 1 to 4.
\item {Semantic Similarity:} Measures sentence-level semantic fidelity by computing the cosine similarity between embeddings of the reconstructed and reference sentences, generated by a pretrained Sentence-BERT model~\cite{emnlp19/nils}.
\end{itemize}

\subsection{Reconstruction Performance over an AWGN Channel}
Fig.\,\ref{fig:awgn snr img} presents the image reconstruction performance in terms of PSNR, SSIM, and LPIPS over an AWGN channel. Conventional schemes exhibit a pronounced cliff effect: JPEG remains near \(11.4\)\,dB up to \(6\)\,dB SNR and recovers to about \(20.7\)\,dB only at \(8\)\,dB and above, while BPG jumps from about \(11.5\)\,dB at \(6\)\,dB to \(25.4\)\,dB at \(8\)\,dB. This degradation stems from the inability of channel coding to correct bit errors below a critical threshold and the subsequent error propagation during entropy decoding. In contrast, learning-based methods vary more smoothly with SNR. They outperform JPEG at every tested SNR, and MSC matches or exceeds BPG once the SSCC cliff has passed. Specifically, MSC surpasses SwinJSCC by \(1.34\)\,dB in PSNR and \(9.02\%\) in SSIM, while reducing LPIPS by \(7.72\%\) at \(0\)\,dB SNR. The PSNR gap over SwinJSCC grows to \(1.85\)\,dB at \(12\)\,dB, where MSC reaches \(26.75\)\,dB. U-DeepSC saturates near \(17.37\)\,dB PSNR at high SNR, below both JPEG and BPG. Its transmitted image features are fully quantized by a discrete codebook, which caps the expressiveness available for fine visual detail, and the resulting indices are sent digitally, so residual channel errors corrupt the recovered codewords rather than being absorbed as soft analog distortion. MSC instead quantizes only the shared branch and keeps a continuous unique-feature path for image details. For images, most channel uses are allocated to analog unique features, so the shared indices mainly provide compact cross-modal semantic anchors rather than carrying most of the image information.

Fig.\,\ref{fig:awgn snr tex} illustrates the text transmission performance versus SNR. Huffman and ZSTD coding recover losslessly at high SNR, but collapse at low SNR, where the WER reaches \(1\) and the BLEU score is \(0\). SemCom systems avoid this cliff and remain usable when conventional codecs fail. In the proposed MSC, COCO captions largely overlap with visual content, so most caption semantics can be represented by shared codebook indices with a short unique residual. This explains the low WER and high BLEU once indices are decoded correctly, which reflects in-domain caption recovery rather than open-domain bit-exact text transmission. At \(0\)\,dB SNR, MSC is slightly weaker than DeepSC because digital index decoding can still fail. As the SNR increases and the index link becomes reliable, MSC surpasses DeepSC on all three metrics, showing the benefit of the hybrid digital-analog design. At \(12\)\,dB, its WER and BLEU are \(0.0288\) and \(0.9382\), versus \(0.1930\) and \(0.7624\) for DeepSC. U-DeepSC remains weaker on text for the same two reasons as on images: codebook quantization limits how precisely word-level detail can be represented, and digital index transmission cannot gracefully absorb channel noise the way a continuous unique stream can.

\begin{figure*}[t]
    \centering
    \captionsetup[subfigure]{labelformat=empty}
    \subfloat[\label{fig.eva snr PSNR}]{
	\includegraphics[scale =0.36]{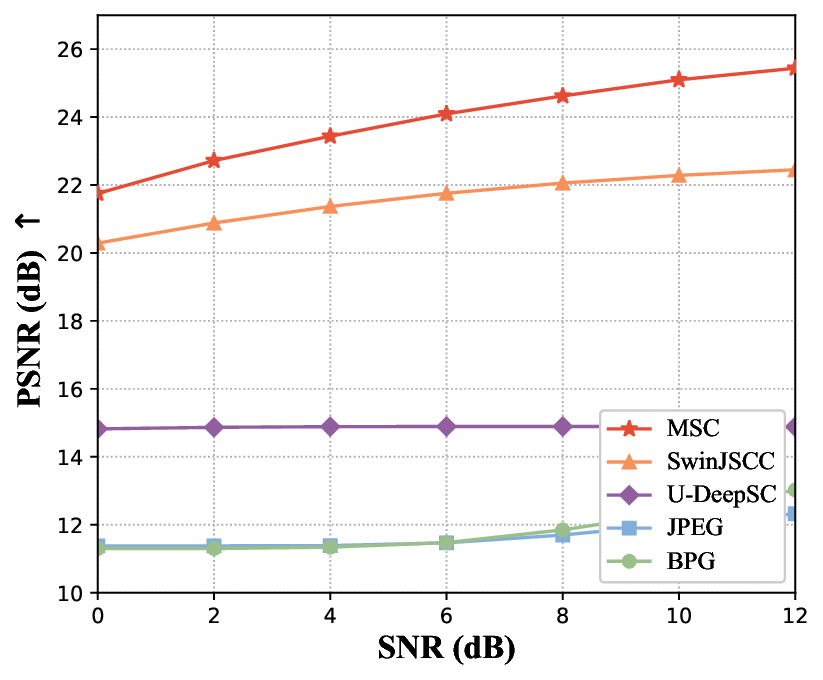}} 
    % \hfill 
    \subfloat[\label{fig.eva snr SSIM}]{
        \includegraphics[scale =0.36]{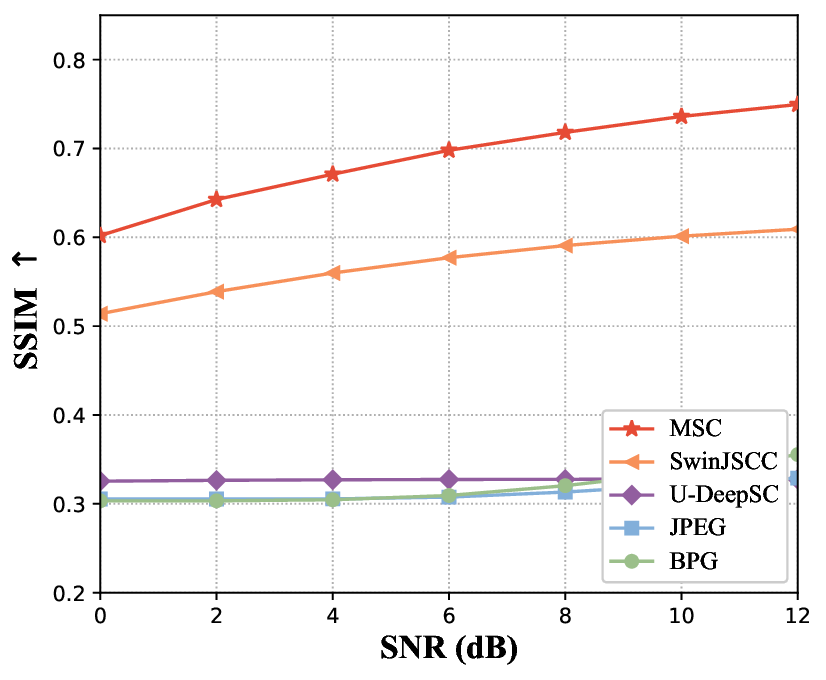}}
    % \hfill 
    \subfloat[\label{fig.eva snr LPIPS}]{
        \includegraphics[scale =0.36]{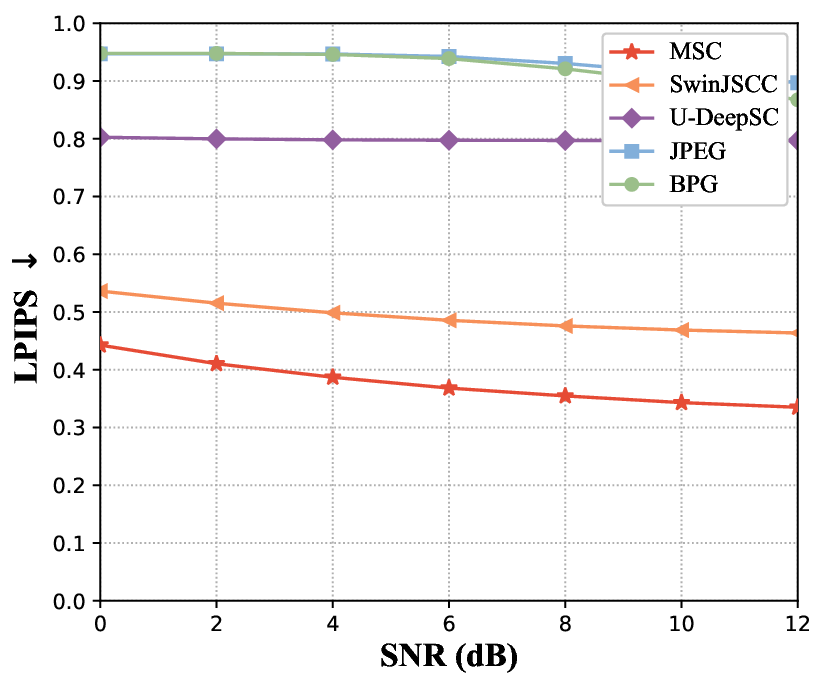}}
    \vspace{-0.5cm}
    \caption{The performance of image transmission versus SNR over an EVA channel with $\iota=0.92$, evaluated in terms of (a) PSNR, (b) SSIM, and (c) LPIPS.}
    \label{fig:eva snr img}
    \vspace{-0.4cm}
\end{figure*}

\begin{figure*}[t]
    \centering
    \captionsetup[subfigure]{labelformat=empty}
    \subfloat[\label{fig.eva snr WER}]{
	\includegraphics[scale =0.36]{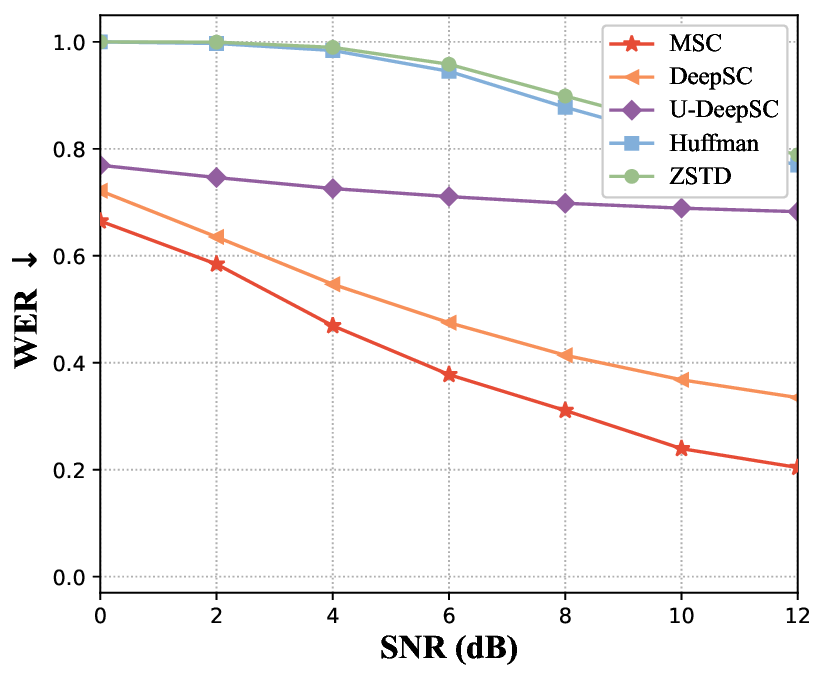}} 
    % \hfill 
    \subfloat[\label{fig.eva snr BLEU}]{
        \includegraphics[scale =0.36]{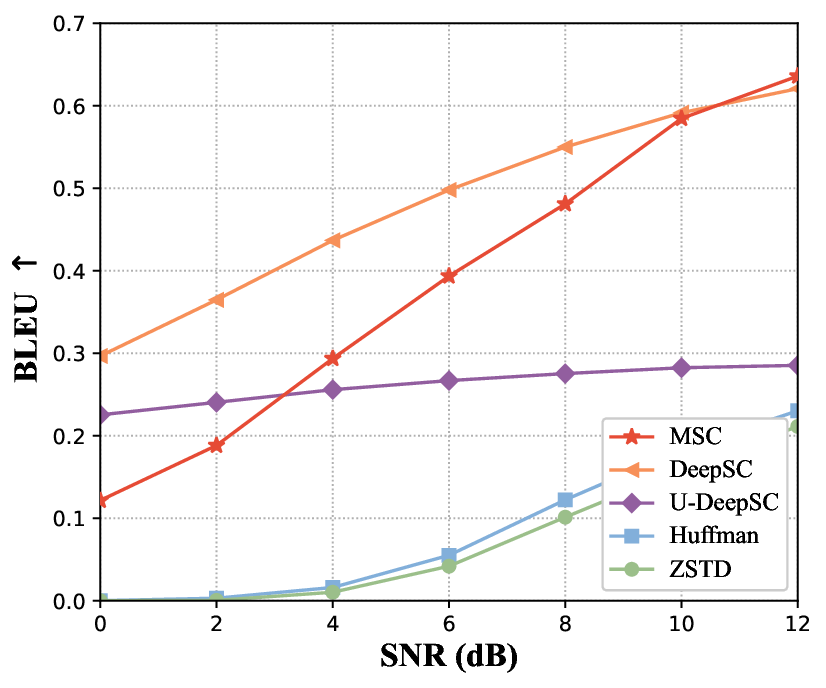}}
    % \hfill 
    \subfloat[\label{fig.eva snr SS}]{
        \includegraphics[scale =0.36]{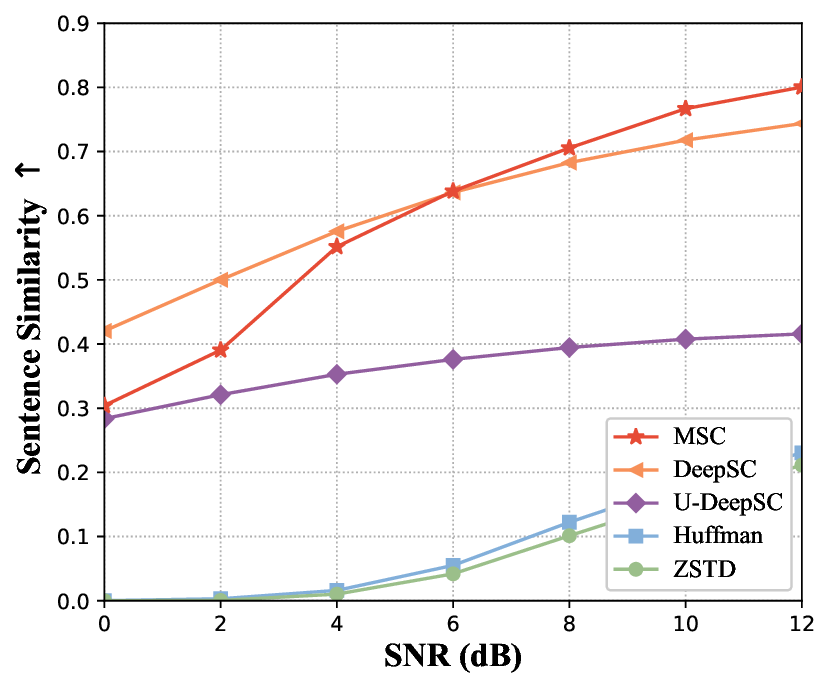}}
    \vspace{-0.5cm}
    \caption{The performance of text transmission versus SNR over an EVA channel with $\iota=0.92$, evaluated in terms of (a) WER, (b) BLEU, and (c) sentence similarity.}
    \label{fig:eva snr tex}
    \vspace{-0.2cm}
\end{figure*}

\subsection{Reconstruction Performance over an EVA Channel}
Figs.\,\ref{fig:eva snr img} and \ref{fig:eva snr tex} report image and text reconstruction over an EVA channel under imperfect CSI with \(\iota=0.92\). At fixed \(\iota\), raising the SNR reduces additive noise but does not remove the mismatch between \(H_{k,t}\) and \(\hat{H}_{k,t}\) in \eqref{eq.channel_factor}. Conventional codecs therefore remain limited: JPEG and BPG reach only \(12.32\) and \(13.02\)\,dB PSNR at \(12\)\,dB SNR, and Huffman and ZSTD stay far from lossless recovery, with WERs of \(0.7698\) and \(0.7888\). Learning-based schemes are more robust under this residual estimation error. MSC leads on all three image metrics. Its PSNR rises from \(21.75\) to \(25.44\)\,dB as the SNR increases from \(0\) to \(12\)\,dB, and its advantage over SwinJSCC grows from \(1.46\) to \(2.99\)\,dB. At \(12\)\,dB, its SSIM and LPIPS are \(0.7495\) and \(0.3349\), compared with \(0.6091\) and \(0.4635\) for SwinJSCC. U-DeepSC stays near \(14.9\)\,dB PSNR across the tested range, again limited by codebook quantization and digital index transmission, which are further stressed by fading and CSI mismatch. On text, MSC has the lowest WER at every tested SNR and reaches \(0.2041\) at \(12\)\,dB, compared with \(0.3348\) for DeepSC. Consistent with the AWGN case, MSC is weaker than DeepSC on BLEU and sentence similarity at low SNR, where shared-index decoding can still fail, then improves as the index link becomes reliable. Sentence similarity exceeds DeepSC from \(6\)\,dB onward and reaches \(0.8004\) at \(12\)\,dB, compared with \(0.7437\). BLEU exceeds DeepSC only at \(12\)\,dB, with a \(2.45\%\) relative gain. Training with channel-estimation mismatch further strengthens MSC under imperfect CSI.

\begin{table*}[t]
    \centering
    \caption{Average Number of Complex Symbols and Inference Complexity per Image and Sentence}
    \label{tab:comparison}
    {\color{black}
    \begin{tabular}{lccccc}
      \toprule
      \multicolumn{6}{c}{\textbf{Image Transmission}} \\
      \cmidrule(r){1-6}
      \textbf{Method}      & \textbf{JPEG} & \textbf{BPG} & \textbf{SwinJSCC} & \textbf{U-DeepSC} & \textbf{MSC} \\
      \midrule
      Average Symbols      & 7399.0        & 	
      4323.7       & 4096.0            & 3920.0            & 4136.0       \\
      Complexity (GFLOPs)  & -         & -       & 5.22               & 4.06               & 11.00        \\
      \midrule
      \multicolumn{6}{c}{\textbf{Text Transmission}} \\
      \cmidrule(r){1-6}
      \textbf{Method}      & \textbf{Huffman} & \textbf{ZSTD} & \textbf{DeepSC} & \textbf{U-DeepSC} & \textbf{MSC} \\
      \midrule
      Average Symbols      & 105.7            & 126.7         & 100.1           & 89.1              & 82.0         \\
      Complexity (GFLOPs)  & -          & -        & 0.25            & 3.20              & 0.51         \\
      \bottomrule
    \end{tabular}}
    \vspace{1mm}
\end{table*}

\subsection{Communication Overhead and Computational Complexity}
Table\,\ref{tab:comparison} reports the average number of complex symbols per image and per sentence, including the shared-index overhead of MSC. The shared indices occupy \(N_c\times\log_2 M=4\times\log_2 1024=40\) bits. With rate-\(1/2\) LDPC coding and QPSK, these bits map to \(40\) complex symbols and are added to the analog unique-feature symbols. Directly sending the continuous shared features \(\mathbf{s}\in\mathbb{R}^{L_s\times D_s}\) as \(32\)-bit floats would instead require \(32\,L_s D_s=4096\) bits before channel coding, since the codebook entries are not energy-normalized for analog packing. The index interface therefore replaces a high-rate digital payload on the shared branch with only \(40\) coded QPSK symbols. On images, MSC uses \(4136.0\) complex symbols and stays close to SwinJSCC and U-DeepSC, while remaining far below JPEG. On text, MSC uses the fewest symbols among all compared schemes. The same \(40\) index symbols form about half of the text budget and about \(1\%\) of the image budget, in line with the greater weight of shared semantics in captions.

The same table lists inference complexity in GFLOPs for the neural methods, and the conventional codecs are entered as $-$. One image costs \(11.00\) GFLOPs with MSC, higher than SwinJSCC and U-DeepSC, because the image encoder uses four Transformer blocks of dimension \(384\) on the full patch sequence and the unified decoder adds four further blocks. One sentence costs \(0.51\) GFLOPs, above DeepSC and well below U-DeepSC. The text encoder has two Transformer blocks and a short token sequence, which keeps the text cost below the image path.

\begin{figure}[t]
    \centering
    \subfloat[Image LPIPS]{\includegraphics[width=0.7\columnwidth]{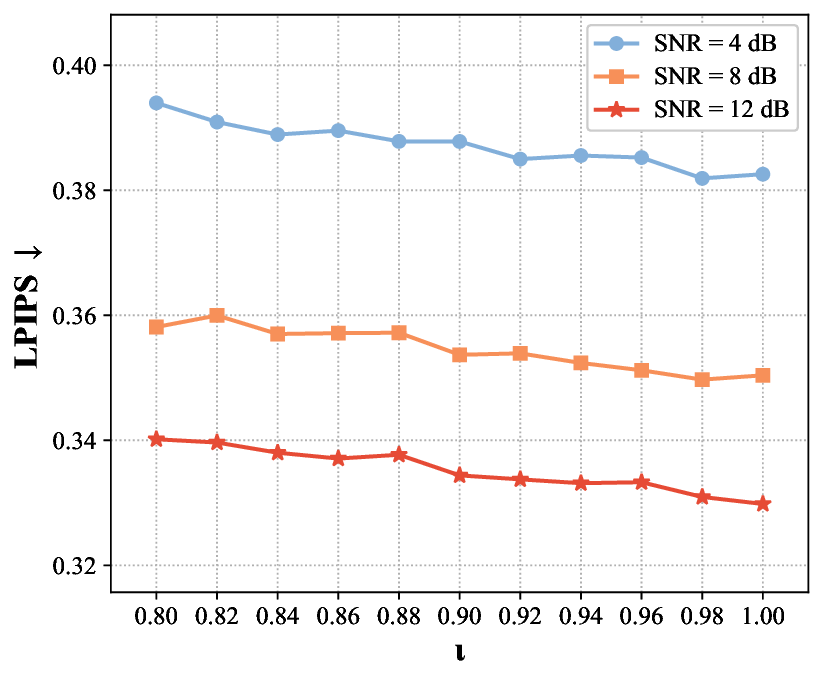}
    \label{fig.eva iota LPIPS}}\\
    \subfloat[Text sentence similarity]{\includegraphics[width=0.7\columnwidth]{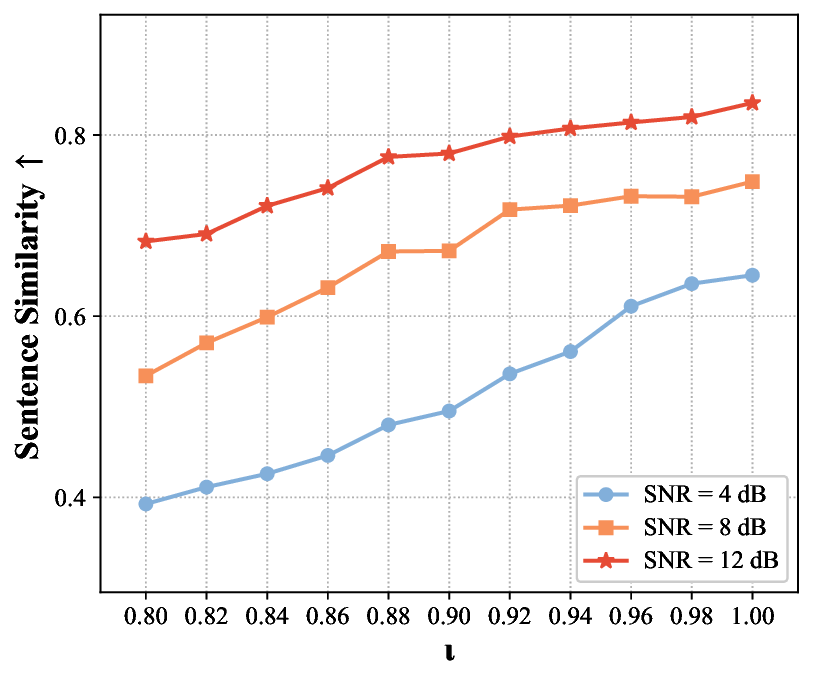}
    \label{fig.eva iota SS}}
    \caption{Reconstruction of MSC versus the CSI correlation coefficient $\iota$ over an EVA channel. The curves are evaluated at SNR values of $4$, $8$, and $12$ dB. (a) Image LPIPS. (b) Text sentence similarity.}
    \label{fig:eva iota}
    \vspace{-0.2cm}
\end{figure}

\begin{table*}[t]
    \centering
    \caption{Ablation Results over an AWGN Channel at $12$\,dB. SS Denotes Sentence Similarity, and MR Denotes the Median Rank of the Paired Sample}
    \label{tab:ablation}
    \begin{tabular}{lcccccccc}
      \toprule
      \multirow{2}{*}{\textbf{Model}} & \multicolumn{2}{c}{\textbf{Normal Transmission}} & \multicolumn{2}{c}{\textbf{Index Only}} & \multicolumn{2}{c}{\textbf{Shared Swap}} & \multicolumn{2}{c}{\textbf{Retrieval}} \\
      \cmidrule(lr){2-3} \cmidrule(lr){4-5} \cmidrule(lr){6-7} \cmidrule(lr){8-9}
                  & PSNR (dB) & SS    & BLEU  & SS    & PSNR (dB) & SS    & Shared R@10 (\%) & Unique MR \\
      \midrule
      MSC         & 26.75     & 0.967 & 0.116 & 0.593 & 26.37     & 0.635 & 31.0             & 2501      \\
      w/o KD      & 26.14     & 0.956 & 0.058 & 0.509 & 25.59     & 0.570 & 30.0             & 2496      \\
      w/o InfoNCE & 26.93     & 0.961 & 0.064 & 0.373 & 24.13     & 0.177 & 0.0              & 2528      \\
      w/o vCLUB   & 26.29     & 0.933 & 0.030 & 0.365 & 25.96     & 0.621 & 31.4             & 1543      \\
      \bottomrule
    \end{tabular}
    \vspace{-0.3cm}
  \end{table*}

\subsection{Robustness to CSI Correlation}
Fig.\,\ref{fig:eva iota} tests the EVA-trained MSC when the CSI correlation at evaluation differs from the training value \(\iota=0.92\). The test coefficient is swept from \(0.80\) to \(1.00\) in steps of \(0.02\), at SNR values of \(4\), \(8\), and \(12\)\,dB. A smaller \(\iota\) increases the residual after zero-forcing equalization in \eqref{eq.channel_factor} and raises the chance that a shared index is decoded incorrectly and replaced by the learned erasure token.

On all three SNR curves, image LPIPS decreases gradually as \(\iota\) increases, while text sentence similarity rises more sharply over the same sweep. This gap follows from the hybrid design: image recovery relies mainly on analog unique features, which degrade smoothly under residual CSI error, whereas caption recovery depends more on correctly decoded shared indices, so digital index errors hurt text more as \(\iota\) falls. The \(12\)\,dB curve keeps the lowest LPIPS and the highest sentence similarity, followed by the \(8\)\,dB and \(4\)\,dB curves. The training operating point \(\iota=0.92\) lies between the two ends of the sweep, including perfect CSI at \(\iota=1\), and yields values close to the high-\(\iota\) regime. MSC therefore remains effective when the test CSI correlation departs from the value seen in training.

\subsection{Ablation Study}
\label{Sec.Ablation}
To isolate the roles of alignment, decoupling, and knowledge distillation, we construct three variants of MSC: w/o InfoNCE, w/o vCLUB, and w/o knowledge distillation. The variants are trained and tested under the same setting as MSC over an AWGN channel at $12$\,dB. Besides normal transmission, Table\,\ref{tab:ablation} reports three tests on the shared branch. (i) Index only: the receiver decodes from the received shared indices, and the received unique features are replaced by Gaussian noise of the same power. (ii) Shared swap: the received shared codewords of the image and text are exchanged before decoding, and each modality keeps its own unique features. (iii) Retrieval: cross-modal retrieval is performed using cosine similarity of the mean-pooled shared features before quantization and of the mean-pooled unique features. Shared-feature Recall@10 (Shared R@10) is averaged over image-to-text and text-to-image retrieval and reflects the alignment of shared features. For the unique features, random guessing gives a median rank (MR) of about $2500$, so a lower Unique MR indicates weaker decoupling.
  
Under normal transmission, w/o InfoNCE reaches the highest PSNR of $26.93$\,dB, while w/o knowledge distillation and w/o vCLUB fall below MSC in both PSNR and sentence similarity. Unique features can still carry much of the reconstruction load and partially compensate for a weaker shared branch. The remaining tests therefore remove or replace the unique features to expose each loss term.
  
InfoNCE aligns shared features across modalities by attracting matched pairs and pushing apart unmatched pairs in the shared embedding space. Without it, Shared R@10 drops from $31.0\%$ to $0\%$, and cross-modal retrieval fails. Shared-swap sentence similarity falls from $0.635$ to $0.177$, so the shared codewords of one modality cannot replace those of the other. Index-only sentence similarity also drops to $0.373$. The Shared R@10 of w/o vCLUB and w/o knowledge distillation stays near that of MSC, since this metric is governed mainly by InfoNCE.

vCLUB promotes the shared--unique decomposition by reducing dependence between the two branches. Without it, Unique MR decreases from $2501$ to $1543$, whereas the other models remain near the random level of $2500$. Shared content then overlaps with the unique features, which lowers the reconstruction ceiling under normal transmission: sentence similarity drops to $0.933$ and PSNR to $26.29$\,dB. Index-only recovery is the weakest among all variants, with sentence similarity $0.365$ and BLEU $0.030$.

\begin{figure}[t]
    \centering
    \subfloat[Shared features]{\includegraphics[width=0.7\columnwidth]{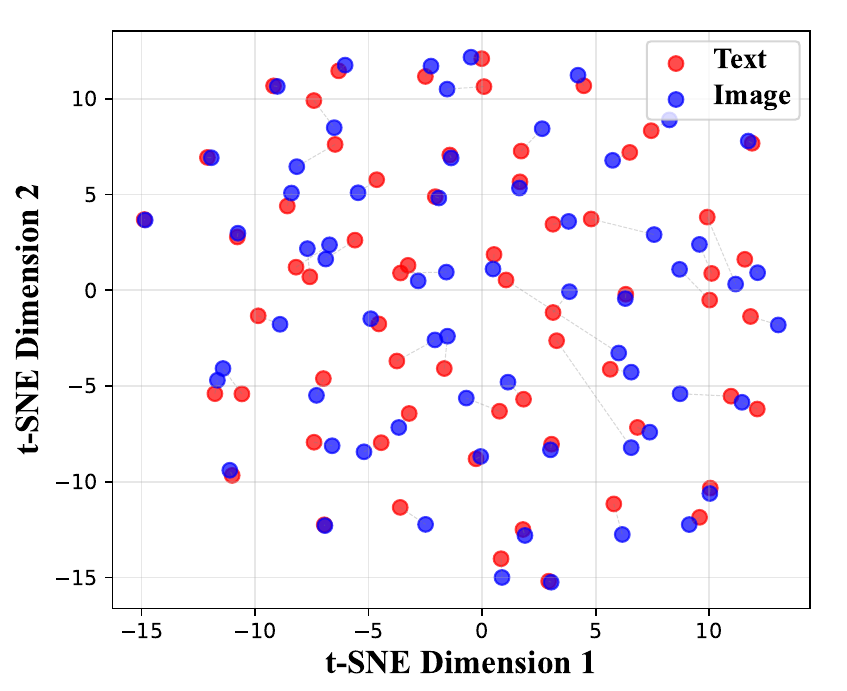}
    \label{fig:tsne_shared}}\\
    \subfloat[Unique features]{\includegraphics[width=0.7\columnwidth]{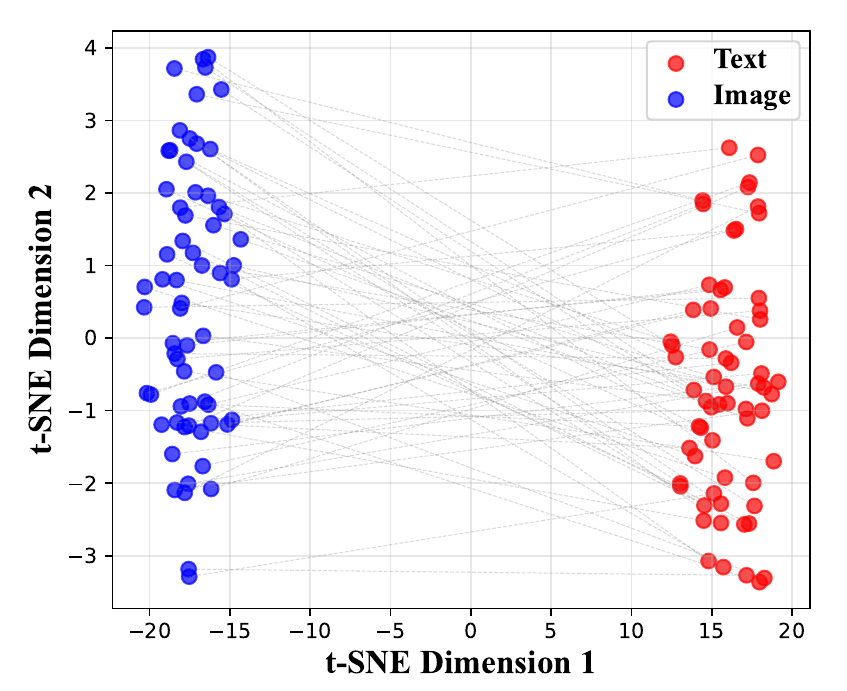}
    \label{fig:tsne_unique}}
    \caption{The t-SNE visualizations of the learned features: (a) shared feature space and (b) unique feature space. Red denotes text, blue denotes images, and paired samples are connected by gray dashed lines.}
    \label{fig:tsne_combined}
    \vspace{-0.2cm}
\end{figure}

Knowledge distillation improves the semantic content of the shared indices for recovery. Removing it reduces normal-transmission PSNR from $26.75$\,dB to $26.14$\,dB and sentence similarity from $0.967$ to $0.956$. The degradation is clearer when decoding relies on the shared branch: index-only sentence similarity falls from $0.593$ to $0.509$ and BLEU is halved from $0.116$ to $0.058$, while shared-swap PSNR and sentence similarity also decline. Shared R@10 remains close to that of MSC, so knowledge distillation mainly strengthens the usefulness of the shared indices rather than cross-modal interchangeability beyond InfoNCE. In summary, InfoNCE aligns the shared features, vCLUB decouples them from the unique features, and knowledge distillation enriches the shared indices. Only the complete MSC maintains strong recovery when the unique features are replaced by noise or when the shared codewords are swapped between modalities.

\subsection{Alignment and Decoupling Visualization}
Fig.\,\ref{fig:tsne_combined} visualizes the learned representations with t-distributed stochastic neighbor embedding (t-SNE). Fig.\,\ref{fig:tsne_shared} and Fig.\,\ref{fig:tsne_unique} show the shared and unique feature spaces, respectively. Text and image samples are marked in red and blue, and paired samples are linked by gray dashed lines. In the shared space, paired cross-modal samples form tight clusters, which indicates that the shared features capture modality-invariant semantics. In the unique space, text and image samples separate clearly while remaining discriminative within each modality, which is consistent with the preservation of modality-specific information. These plots therefore support the intended decomposition: shared features are aligned across modalities, and unique features retain modality-specific content.

\begin{figure}
    \centering
    \includegraphics[width=0.35\textwidth]{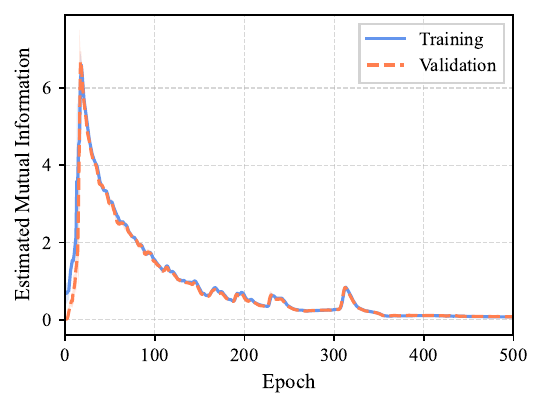}
    \vspace{-0.2cm}
    \caption{Estimated mutual information between shared and unique features.}
    \label{fig:mutual_info}
    \vspace{-0.2cm}
\end{figure}

Fig.\,\ref{fig:mutual_info} further tracks the estimated mutual information between shared and unique features during training. The curve first rises as the mutual information estimator learns the dependency, then declines and stabilizes near zero. This trend shows that the optimization reduces the statistical dependence between the two branches, i.e., the decoupling term in \eqref{eq.leak_shannon} is driven close to zero. Together with cross-modal alignment, this keeps the leakage bound in \textbf{Proposition}\,\ref{prop:link} small.

\section{Conclusion}
\label{Sec.Conclusion}
This paper has presented MSC, a multimodal SemCom framework that decomposes each modality into shared and unique representations for efficient transmission and source recovery. Cross-modal alignment and shared--unique decoupling are encouraged through InfoNCE and vCLUB, while a shared knowledge base maps the shared components to compact indices for digital transmission and leaves the unique components to analog JSCC. Knowledge distillation from a vision foundation model further enriches the quantized shared image features during training. Experiments on image--text pairs over AWGN and EVA channels show that MSC improves reconstruction over SSCC and representative SemCom baselines at a matched communication cost. Additional evaluations indicate gradual degradation under CSI-correlation mismatch, complementary roles of the alignment, decoupling, and knowledge-distillation terms, and clear separation between aligned shared features and modality-specific unique features.

\appendices
\section{Mathematical Derivation of Multimodal Semantic Representation Gap}
\label{App:Semantic_Theory}
This appendix derives the semantic joint entropy \(H_s(\cdot,\cdot)\) and semantic mutual information \(I_s(\cdot;\cdot)\) used in \textbf{Definition}\,\ref{def:alignment} and \textbf{Definition}\,\ref{def:gap}, by extending the synonymous mapping~\cite{sbcs25/kai} to two modalities.
Let \(\mathcal{X}\) and \(\mathcal{Y}\) denote the syntactic sets of observable samples, and let \(\tilde{\mathcal{X}}\) and \(\tilde{\mathcal{Y}}\) denote the corresponding sets of discrete semantic concepts. For the entropy calculations below, we consider discrete or digitized source samples and assume that the relevant entropies are finite. Synonymous mappings associate each concept with a set of samples that share the same meaning. We write these as \(f_x:\tilde{\mathcal{X}}\to\mathcal{X}\) and \(f_y:\tilde{\mathcal{Y}}\to\mathcal{Y}\), which partition the syntactic sets into disjoint equivalence classes \(\{\mathcal{X}_{i}\}\) and \(\{\mathcal{Y}_{j}\}\). For any concept \(\tilde{x}_{i}\in\tilde{\mathcal{X}}\), the synonymous set is \(\mathcal{X}_{i}\subset\mathcal{X}\). Because the synonymous sets form partitions of the syntactic sets, every sample belongs to exactly one synonymous set. This defines the deterministic maps \(g_x(x)=\tilde{x}_{i}\) for \(x\in\mathcal{X}_{i}\) and \(g_y(y)=\tilde{y}_{j}\) for \(y\in\mathcal{Y}_{j}\), so that \(\tilde{\mathbf{X}}=g_x(\mathbf{X})\) and \(\tilde{\mathbf{Y}}=g_y(\mathbf{Y})\).
Let \(C_{ij}=(\tilde{x}_{i},\tilde{y}_{j})\) denote a joint semantic event on the pair set \(\mathcal{X}_{i}\times\mathcal{Y}_{j}\), with
\begin{equation}
    p(C_{ij}) = \sum_{x \in \mathcal{X}_{i}} \sum_{y \in \mathcal{Y}_{j}} p(x, y).
\end{equation}
The marginals are \(p(\tilde{x}_{i})=\sum_{j}p(C_{ij})\) and \(p(\tilde{y}_{j})=\sum_{i}p(C_{ij})\).
Following~\cite{sbcs25/kai}, the semantic joint entropy \(H_s\) and semantic mutual information \(I_s\) are given by
\begin{equation}
    H_s(\tilde{\mathbf{X}}, \tilde{\mathbf{Y}})
    = -\sum_{i}\sum_{j} p(C_{ij})\log p(C_{ij}),
\end{equation}
\begin{equation}
    I_s(\tilde{\mathbf{X}}; \tilde{\mathbf{Y}})
    = \sum_{i}\sum_{j} p(C_{ij})
    \log \frac{p(C_{ij})}{p(\tilde{x}_{i})p(\tilde{y}_{j})},
\end{equation}
respectively.
Their difference equals the sum of conditional semantic entropies:
\begin{equation}
    \begin{aligned}
        &H_s(\tilde{\mathbf{X}}, \tilde{\mathbf{Y}}) - I_s(\tilde{\mathbf{X}}; \tilde{\mathbf{Y}}) \\
        &= -\sum_{i,j} p(C_{ij}) \log p(C_{ij}) - \sum_{i,j} p(C_{ij}) \log \frac{p(C_{ij})}{p(\tilde{x}_{i}) p(\tilde{y}_{j})} \\
        &= \sum_{i,j} p(C_{ij}) \log \frac{p(\tilde{x}_{i}) p(\tilde{y}_{j})}{p(C_{ij})^2} \\
        &= \sum_{i,j} p(C_{ij}) \log \frac{p(\tilde{x}_{i})}{p(C_{ij})} + \sum_{i,j} p(C_{ij}) \log \frac{p(\tilde{y}_{j})}{p(C_{ij})} \\
        &= H_s(\tilde{\mathbf{Y}} | \tilde{\mathbf{X}}) + H_s(\tilde{\mathbf{X}} | \tilde{\mathbf{Y}}).
    \end{aligned}
\end{equation}
Thus \textbf{Definition}\,\ref{def:alignment} holds if and only if \(H_s(\tilde{\mathbf{Y}}|\tilde{\mathbf{X}})+H_s(\tilde{\mathbf{X}}|\tilde{\mathbf{Y}})=0\).
Since conditional entropy is non-negative, this is equivalent to both conditional entropies being zero.
Equivalently, each semantic concept occurring with nonzero probability in one modality uniquely determines a concept in the other, and vice versa.
\textbf{Definition}\,\ref{def:gap} corresponds to the case \(H_s(\tilde{\mathbf{Y}}|\tilde{\mathbf{X}})+H_s(\tilde{\mathbf{X}}|\tilde{\mathbf{Y}})>0\), meaning that at least one modality contains semantic information that cannot be fully determined from the other.

In Section\,\ref{Sec.System Model}, the shorthand \(H_s(\mathbf{X},\mathbf{Y})\) and \(I_s(\mathbf{X};\mathbf{Y})\) mean that the measures are evaluated on these semantic variables, i.e., \(H_s(\mathbf{X},\mathbf{Y})\triangleq H_s(\tilde{\mathbf{X}},\tilde{\mathbf{Y}})\) and \(I_s(\mathbf{X};\mathbf{Y})\triangleq I_s(\tilde{\mathbf{X}};\tilde{\mathbf{Y}})\). Likewise, \(H_s(\mathbf{X}|\mathbf{Y})\triangleq H_s(\tilde{\mathbf{X}}|\tilde{\mathbf{Y}})\), with the reverse conditional entropy defined analogously.
Since \(\tilde{\mathbf{X}}\) and \(\tilde{\mathbf{Y}}\) are deterministic functions of \(\mathbf{X}\) and \(\mathbf{Y}\), these semantic measures coincide with the classical entropy and mutual information of \((\tilde{\mathbf{X}},\tilde{\mathbf{Y}})\), and \eqref{eq.entropy_split} follows from \(H(\mathbf{X})=H(\tilde{\mathbf{X}})+H(\mathbf{X}|\tilde{\mathbf{X}})\) and \(H(\tilde{\mathbf{X}})=I(\tilde{\mathbf{X}};\tilde{\mathbf{Y}})+H(\tilde{\mathbf{X}}|\tilde{\mathbf{Y}})\).

\section{Proof of Proposition~\ref{prop:link}}
\label{App:Proof_Prop}
Since $\mathbf{S}_x$ and $\mathbf{U}_x$ are functions of $\mathbf{X}$, for any random variable $\mathbf{V}$ that is a function of $\mathbf{Y}$, we have
\begin{align}
    I(\mathbf{U}_x;\mathbf{V}) 
    &\overset{(a)}{\le} I(\mathbf{U}_x;\mathbf{S}_x,\mathbf{V}) 
    = I(\mathbf{U}_x;\mathbf{S}_x) + I(\mathbf{U}_x;\mathbf{V}|\mathbf{S}_x) \notag \\
    &\overset{(b)}{\le} I(\mathbf{U}_x;\mathbf{S}_x) + I(\mathbf{X};\mathbf{V}|\mathbf{S}_x) \notag \\
    &\overset{(c)}{=} I(\mathbf{U}_x;\mathbf{S}_x) + I(\mathbf{X};\mathbf{V}) - I(\mathbf{S}_x;\mathbf{V}),
    \label{eq.proof_core}
\end{align}
where (a) follows from the chain rule and the non-negativity of mutual information; (b) holds because $\mathbf{U}_x$ is a function of $\mathbf{X}$, so that $I(\mathbf{X};\mathbf{V}|\mathbf{S}_x)=I(\mathbf{X},\mathbf{U}_x;\mathbf{V}|\mathbf{S}_x)\ge I(\mathbf{U}_x;\mathbf{V}|\mathbf{S}_x)$; and (c) holds because $\mathbf{S}_x$ is a function of $\mathbf{X}$, so that $I(\mathbf{X};\mathbf{V})=I(\mathbf{X},\mathbf{S}_x;\mathbf{V})=I(\mathbf{S}_x;\mathbf{V})+I(\mathbf{X};\mathbf{V}|\mathbf{S}_x)$.

\textit{Proof of \eqref{eq.leak_shannon}:} Set $\mathbf{V}=\mathbf{Y}$ in \eqref{eq.proof_core}. Note that $I(\mathbf{U}_x;\mathbf{S}_x)\le I(\mathbf{S}_x,\mathbf{S}_y;\mathbf{U}_x,\mathbf{U}_y)$ by the chain rule, and $I(\mathbf{S}_x;\mathbf{Y})\ge I(\mathbf{S}_x;\mathbf{S}_y)$ by the data-processing inequality, since $\mathbf{S}_y$ is a function of $\mathbf{Y}$.

\textit{Proof of \eqref{eq.align_ceiling}:} Under condition (\romannumeral1), $\mathbf{S}_x\to\tilde{\mathbf{X}}\to\tilde{\mathbf{Y}}\to\mathbf{S}_y$ forms a Markov chain. The data-processing inequality gives $I(\mathbf{S}_x;\mathbf{S}_y)\le I(\tilde{\mathbf{X}};\tilde{\mathbf{Y}})=I_s(\mathbf{X};\mathbf{Y})$.

\textit{Proof of \eqref{eq.leak_semantic}:} Set $\mathbf{V}=\tilde{\mathbf{Y}}$ in \eqref{eq.proof_core}. Under condition (\romannumeral2), $I(\mathbf{X};\tilde{\mathbf{Y}})=I(\tilde{\mathbf{X}};\tilde{\mathbf{Y}})+I(\mathbf{X};\tilde{\mathbf{Y}}|\tilde{\mathbf{X}})=I_s(\mathbf{X};\mathbf{Y})$. Under condition (\romannumeral1), $\mathbf{S}_y$ is a function of $\tilde{\mathbf{Y}}$, and thus $I(\mathbf{S}_x;\tilde{\mathbf{Y}})\ge I(\mathbf{S}_x;\mathbf{S}_y)$. Combining these with $I(\mathbf{U}_x;\mathbf{S}_x)\le I(\mathbf{S}_x,\mathbf{S}_y;\mathbf{U}_x,\mathbf{U}_y)$ completes the proof. The bounds for $\mathbf{U}_y$ follow by exchanging the roles of the two modalities.

\bibliographystyle{IEEEtran}
% \balance
\bibliography{References}

\end{document}